\documentclass[twocolumn,aps,prc,superscriptaddress,floatfix]{revtex4-1}

\newcommand{\tsf}{t_\textup{sf}}
\newcommand{\ezpe}{\epsilon_\mathrm{vib}}
\newcommand{\BI}{{B_{I}}}
\newcommand{\BII}{{B_{II}}}
\newcommand{\EII}{{E_{II}}}
\usepackage{graphicx}
\usepackage{amsmath}
\usepackage{amssymb}
\usepackage{bm}
\usepackage[colorlinks,citecolor=blue]{hyperref}
\usepackage[colorinlistoftodos]{todonotes}
\usepackage[caption=false]{subfig}
\graphicspath{{images/}}

\begin{document}

\title{Fission properties of superheavy nuclei for r-process calculations}%

\author{Samuel~A.~Giuliani}%
\email{giuliani@nscl.msu.edu}
\altaffiliation[Present address: ]{NSCL/FRIB Laboratory, 
Michigan State University, East Lansing, Michigan 48824, USA.}
\affiliation{Institut f{\"u}r Kernphysik (Theoriezentrum), Technische
  Universit{\"a}t Darmstadt, Schlossgartenstra{\ss}e 2, 64289
  Darmstadt, Germany}
\author{Gabriel~Mart\'inez-Pinedo}%
\email{g.martinez@gsi.de}
\affiliation{GSI Helmholtzzentrum f{\"u}r Schwerioneneforschung,
  Planckstra{\ss}e 1, 64291 Darmstadt, Germany} 
\affiliation{Institut f{\"u}r Kernphysik (Theoriezentrum), Technische
  Universit{\"a}t Darmstadt, Schlossgartenstra{\ss}e 2, 64289
  Darmstadt, Germany} 
\author{Luis~M.~Robledo}%
\email{luis.robledo@uam.es}
\affiliation{Departamento de F\'isica Te\'orica, Universidad Aut\'onoma de Madrid, 
E-28049 Madrid, Spain}
\date{\today}%

\begin{abstract}
  We computed a new set of static fission properties suited for
  $r$-process calculations. The potential energy surfaces and
  collective inertias of 3640 nuclei in the superheavy region are
  obtained from Self-Consistent Mean-Field calculations using the
  Barcelona-Catania-Paris-Madrid energy density functional. The
  fission path is computed as a function of the quadrupole moment by
  minimizing the potential energy and exploring octupole and
  hexadecapole deformations. The spontaneous fission lifetimes are
  evaluated employing different schemes for the collective inertias
  and vibrational energy corrections.  This allows to explore the
  sensitivity of the lifetimes to those quantities together with the
  collective ground state energy along the superheavy landscape. We
  computed neutron induced stellar reaction rates relevant for
  $r$-process nucleosynthesis using the Hauser-Feshbach statistical
  approach and study the impact of collective inertias. The
  competition between different reaction channels including neutron
  induced rates, spontaneous fission and alpha decay is discussed for
  typical $r$-process conditions.
\end{abstract}

\maketitle

\section{Introduction}

The theoretical description of the fission process is one of the most
challenging and fascinating problems in nuclear physics. The competition
between the long-range Coulomb repulsion and the short-range strong interaction
drives the evolution of the nucleus from the ground state to the scission
point. Furthermore, they produce quantum mechanical shell effects that allow
for the stability of superheavy elements (SH) in the heaviest regions of the
nuclear landscape. Arguably one of the most interesting applications of
fission concerns the $r$-process nucleosynthesis of superheavy elements. In
scenarios with high neutron densities, like for instance the dynamical ejecta
of neutron stars mergers, the competition between rapid neutron captures and
beta decays of seed nuclei leads to the synthesis of superheavy elements. The
$r$-process path proceeds towards regions of unstable nuclei that undergo
fission, recycling the material to lighter fission
products~\cite{Cowan1991,Martinez-Pinedo2007,Goriely2015}. In these conditions,
fission plays a relevant role not only by modifying the final shape of the
$r$-process abundances~\cite{Goriely2013,Eichler2015}, but also in providing a
mechanism to achieve a robust $r$ process~\cite{Mendoza-Temis2014}. Besides
astrophysical applications, the stability of the nucleus against fission is
also crucial for the experimental synthesis of superheavy nuclei achieved in
laboratories all around the world during the past
years~\cite{Stavsetra2009,Dullmann2010,Oganessian2012,Khuyagbaatar2014} and for
energy production in nuclear reactors.

Nowadays, the main nuclear structure models describing the spontaneous
fission (SF) are the microscopic-macroscopic (MicMac) and the self-consistent
mean-field (SCMF) models (see ref.~\cite{Baran2015} for a recent review on
fission properties of SH nuclei covering MicMac and SCMF models). The MicMac
description pioneered the modern modeling of the fission process in the late
sixties and since then several studies have been successfully applied in
systematic calculations of superheavy nuclei. One of the main advantages of
this method is the calculation of the potential energy surface in
multidimensional spaces using up to five collective degrees of freedom for
systematic calculations, providing an accurate description of multiple fission
paths (see, e.g. refs.~\cite{Jachimowicz2015,Moller2015,Jachimowicz2017} for
recent calculations). On the other hand, 20 years ago fission calculations
based on self-consistent methods with effective interactions entered the scene,
proposing an alternative scheme built on a more microscopic approach. Starting
from an effective energy density functional (EDF), the constrained Hartree-Fock
(HF) and Hartree-Fock-Bogoliubov (HFB) theories permit the calculation of the
fission properties rooted on more microscopic input.  In the last years,
several studies explored the capability of the EDF theory to reproduce the
experimental fission data using either 
Skyrme~\cite{Goriely2007,Goriely2009,Pei2009,Sheikh2009,Erler2012a,
Staszczak2013,Sadhukhan2015}, 
Gogny~\cite{Warda2002,Warda:cluster,Delaroche2006,Warda2012,
Rodriguez-Guzman2014,Rodriguez-Guzman2014a,Rodriguez-Guzman2016} or
relativistic interactions~\cite{Lu2014} (see~\cite{Schunck2016} for a complete
review).

The main objective of this paper is to present the fission properties
of $r$-process nuclei obtained with the Barcelona-Catania-Paris-Madrid
(BCPM) EDF~\cite{Baldo2013}.  As already pointed out 
in~\cite{Goriely2015a}, nowadays
only few global calculations suited for $r$-process calculations are
publicly available. This work is designed to provide a new set of
fission properties based in the SCMF model covering the whole
superheavy landscape and including nuclei with an odd number of
protons and/or neutrons. The main advantage of the EDF theory is that
allows the computation of both the potential energy surface and the
collective inertias from a unique and microscopic footing, providing a
robust framework for the calculation of spontaneous fission lifetimes,
fission barrier heights and isomer excitation energies.  The paper is
outlined as follows. In Sec.~\ref{sec:methods} we briefly introduce
the method used in the calculation of the potential energy surface and
spontaneous fission lifetimes. In Sec.~\ref{sec:results} we summarize
the results of our calculations. We start discussing the benchmark of
BCPM against the experimental data in Sec.~\ref{sec:bench}. The
systematic of fission barriers and spontaneous fission lifetimes is
discussed in Sec.~\ref{sec:barriers} and Sec.~\ref{sec:tsf}.  Sec.
\ref{sec:alpha} is devoted to discuss the competition between
$\alpha$-decay and spontaneous fission and in Sec.~\ref{sec:channels}
the discussion is extended to neutron induced rates of relevance to the 
$r$ process.  Finally, in Sec.~\ref{sec:conclusions} we summarize our
results and outline the future perspectives.
 

\section{Methods}\label{sec:methods} 

In this paper fission is described within the Self-Consistent Mean-Field (SCMF)
approach~\cite{Bender2003} following the traditional HFB theory with
constraining operators as described in Ref.~\cite{Giuliani2013}.  For
completeness we will summarize here the general computational scheme and point
out the uncertainties arising from our method. 

In order to reduce the computational cost, axial symmetry has been preserved in
all the calculations. The impact of releasing this restriction has been object
of several recent studies (see
e.g.~\cite{Erler2012a,Staszczak2013,Rodriguez-Guzman2014}), where it has been
shown that triaxiality can reduce the inner fission barrier height of actinides
up to 2--3 MeV. However, we would like to point here that the role of
triaxiality in fission is still the subject of discussion since some recent
calculations showed that axial symmetry can be fully restored in dynamic
calculations of the fission
process~\cite{Delaroche2006,Sadhukhan2014,Zhao2016a}. Moreover, this reduction
of the fission barrier is compensated by an increase of the collective inertias
and therefore the impact of releasing this symmetry is expected to be small in
the calculation of fission lifetimes.

Due to the preservation of axial symmetry, the mean-value of the multipole
operators, $\langle \bm{Q}_{\mu\nu}\rangle=0$ for all $\nu\neq0$. In order to
explore the impact of octupole and hexadecapole deformations, as well as
asymmetric fission, reflection symmetry is allowed to break at any stage of the
calculations. The basis quantum numbers are restricted by the condition:
\begin{equation} 2n_\perp + |m| + \frac{n_z}{q} \le N^{\text{max}}.
\end{equation} All the calculations were carried with $N^{\text{max}}=17$ and
$q=3/2$. The parameter $q$ represents the ratio of the number of quanta along
the z direction to the number of quanta along the perpendicular one. The value
used favors more shells in the $z$ direction as required in fission. Once the
number of quanta in each direction is fixed, the basis only depends on the
oscillator lengths $b_z$ and $b_\perp$. To diminish the impact of the limited
basis size on the binding energies, it is mandatory to carefully optimize the
oscillator length parameters for each value of the constrained quadrupole moment
considered in the calculation~\cite{Arzhanov2016}. The optimization is carried
out automatically for all the nuclei using the gradient method to find the
minimum of the HFB energy as a function of the two oscillator length variables
$b_\perp$ and $b_z$. The gradient of the energy with respect to $b_\perp$ and
$b_z$ is computed numerically using a three point formula for the numerical
derivative.  The HFB equations were solved using a second-order gradient method,
which provides a fast convergence and allows for an arbitrary generalization of
the numbers of constraints~\cite{Rob:grad}. The main advantage of the
computational scheme described above is that it has already been applied to
several fission calculations using either the Gogny or the BCPM
interactions~\cite{Warda2002,Giuliani2013,Rodriguez-%
Guzman2014,Giuliani2014,Rodriguez-Guzman2014a,Rodriguez-Guzman2016}, and its
capability to describe the fission process is well constrained.

We have used the last version of the BCPM functional recently proposed to
describe the physics of finite nuclei~\cite{Baldo2013}. This functional has also
proved to perform well in a series of calculations of fission properties
including inner fission barrier heights, excitation energy of fission isomers,
outer barrier heights, spontaneous fission lifetimes, etc~\cite{Giuliani2013}.
The idea behind the explicit form of the BCPM functional is to use a simple
polynomial in the density to fit the energy per particle in both symmetric and
neutron nuclear matter as obtained with state of the art many body techniques
and realistic nuclear interactions. The polynomial so obtained is used verbatim
in finite nuclei but using the density of the finite nucleus instead. A standard
contact spin-orbit term is added to reproduce magic numbers. Surface effects are
considered by means of a finite range gaussian interaction acting only in the
direct channel.  The Coulomb interaction is taken, as in many other
calculations, exactly in the direct channel. The exchange Coulomb field is
replaced by the Slater approximation and the repulsive contribution of the
Coulomb interaction to the pairing channel is neglected. For the pairing
interaction, a density dependent pairing interaction has been used.  One of the
distinctive characteristics of the functional is its inclusion of the rotational
energy correction (see below) to compute the ground state binding energy. The
rotational correction is also used in the fitting protocol in spite of the fact
that its inclusion produces some artifacts near magic or semimagic nuclei
\cite{Baldo2013}. They are a direct consequence of computing the rotational
correction after variation and disappear if beyond mean field effects like
configuration mixing of quadrupole deformed shapes are taken into account.
Unfortunately, the artifacts in the rotational correction lead to spurious peaks
in both the $S_n$, $S_{2n}$ and related quantities near magic or semimagic
nuclei, which can produce unphysical abundances in $r$-process calculations. For
this reason, we have decided to remove the rotational correction in the
evaluation of one and two neutron separation energies.  The impact of this
prescription in the $S_n$ rms for $Z>84$ is small increasing its value from 0.28
MeV to 0.36 MeV. However, the rotational correction is maintained in fission
barrier calculations as the artifacts do not appear in the relevant regions of
the path to fission.

\subsection{Spontaneous fission lifetimes, fission barriers and collective 
inertias}\label{sec:WKB}

We start defining the fission tunneling probability, $P$, from a state
at excitation energy $E_x$ through a transition state with an
excitation energy $\varepsilon$ on top of the fission barrier as:
\begin{equation}
  \label{eq:ptunneling}
  P(E_x,\varepsilon)=\frac{1}{1+\exp[2S(E_x,\varepsilon)]}\,. 
\end{equation}
$S$ is the integral action computed along the fission path, $L(Q_{20})$,
between the classical turning points $a$ and $b$:
\begin{equation}\label{eq:Sl}
	\begin{split}
  &S_L(E_x,\epsilon)=\\
  &\int_a^b dQ_{20}\,\sqrt{2\mathcal{M}(Q_{20})[\mathcal{V}
  (Q_{20})+\varepsilon-(E_x+E_{\mathrm{GS}})]}\,,
  \end{split}
\end{equation}
where the fission path is obtained by minimizing the effective potential
energy, $\mathcal{V}(Q_{20})$,  (so-called static approach).
In the case of spontaneous fission the excitation energy of the
nucleus corresponds to the so-called collective ground state energy,
$E_0$, and the transition state is taken at zero excitation energy on
top of the barrier. Hence, the spontaneous fission half-life ($\tsf$)
is given by the 
semiclassical Wentzel-Kramers-Brillouin (WKB) theory~\cite{Brack1972}:
\begin{equation}\label{eq:tsf}
  \tsf=\frac{\ln 2}{n P(E_0,0)}=\frac{2.86 \times 10^{-21}\ \mathrm{s}}{P(E_0,0)},
\end{equation}
with $n$ the number of assaults of the nucleus on the fission barrier per unit
time~\cite{Baran1981}. Eq.~(\ref{eq:Sl}) shows  that
the theoretical description of spontaneous fission is based in three main
ingredients: the collective inertias $\mathcal{M}(Q_{20})$, the effective
potential energy $\mathcal{V}(Q_{20})$ and the energy of the collective ground
state $E_0$. The effective potential energy $\mathcal{V}(Q_{20})$ is obtained
by subtracting the vibrational and rotational zero-point energies from the
total HFB energy:
\begin{equation}\label{eq:Veff}
  \mathcal{V}(Q_{20})=E_\mathrm{HFB}(Q_{20})-\epsilon_\mathrm{vib}(Q_{20})-
  \epsilon_\mathrm{rot}(Q_{20})\,.
\end{equation}
The HFB energy $E_\mathrm{HFB}$ is defined as the expectation value of the
Routhian with constraining operators:
\begin{equation}
  \hat{\mathcal{H}}=\hat{\mathcal{H}}_\mathrm{HFB}+\sum_{\nu=1,2}\lambda_\nu
  \hat Q_{\nu0}+\sum_{\tau=p,n}\lambda_\tau\hat N_\tau\,,
\end{equation} 
$\hat{\mathcal{H}}_\mathrm{HFB}$ being the HFB Hamiltonian,
$\hat N_p(\hat N_n)$ the proton (neutron) number operator, $\lambda_i$
the Lagrange multipliers, $Q_{10}$ the center-of-mass constraint
preventing spurious solutions arising from center-of-mass motion
and $\hat{Q}_{20}$:
\begin{equation}
  \label{eq:1}
  \hat{Q}_{20} = \sum_i^A \hat{z}^2_i - \frac{1}{2} (\hat{x}^2_i+\hat{y}^2_i)\,.
\end{equation}
The rotational energy correction $\epsilon_\mathrm{rot}(Q_{20})$ is
related to the restoration of the rotational symmetry and is computed
in terms of the Yoccoz moments of inertia using the phenomenological
approach of Ref.~\cite{Egido2004}. 
This approach includes a correction to account for the approximations
involved in the evaluation of the Yoccoz moment of inertia.
Finally, the vibrational energy
correction $\epsilon_\mathrm{vib}(Q_{20})$ takes into account for
quantal fluctuations in the collective degree of freedom $Q_{20}$.

\begin{figure*}[tb]
 \includegraphics[width=\textwidth]{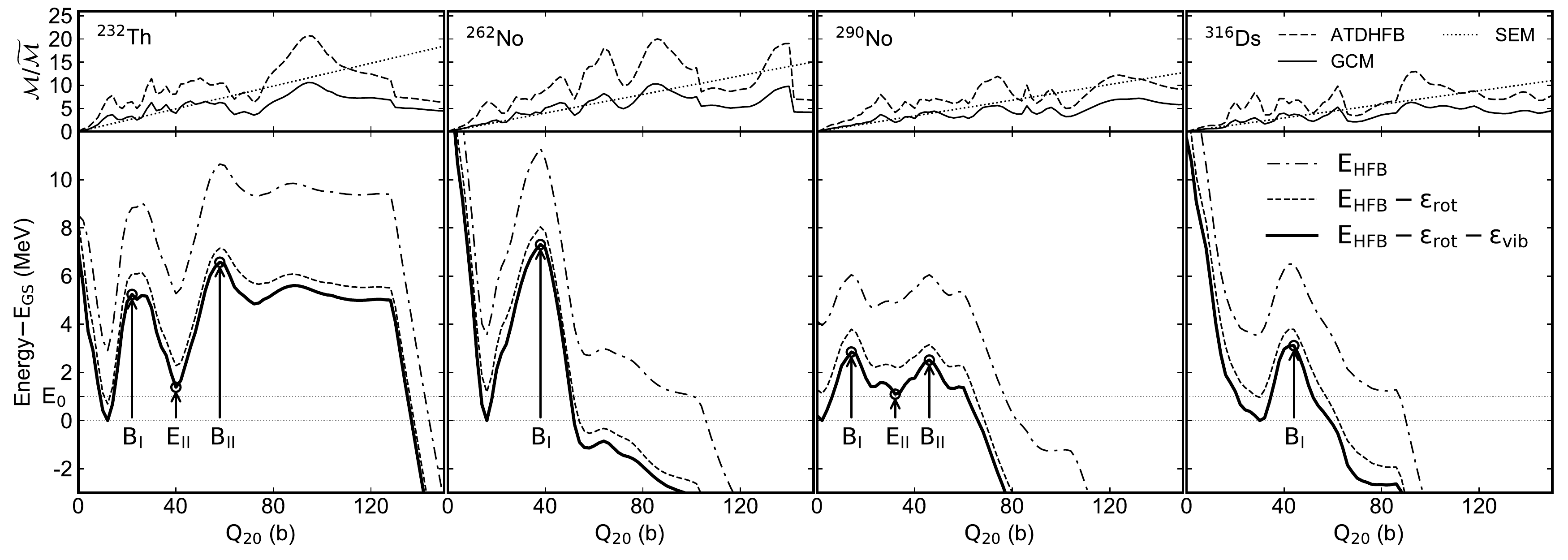}
 \caption{Collective fission properties of $^{232}$Th, $^{262}$No,
   $^{290}$No and $^{316}$Ds as a function of the quadrupole
   moment. Lower panels show different contributions to the potential
   energy surface $\mathcal{V}(Q_{20})$ of Eq.~(\ref{eq:Veff}).  Upper
   panels show collective inertias $\mathcal{M} (Q_{20})$ computed
   with the ATDHFB (dashed line), GCM (solid line) and semiempirical
   inertia formula (dotted line), renormalized to the reduced inertia
   $\widetilde{\mathcal{M}}$.}
 \label{fig:PES}
\end{figure*}

The lower panels of Fig.~\ref{fig:PES} show the different
contributions of Eq.~(\ref{eq:Veff}) to the Potential Energy Surface
(PES) in four different nuclei.  Clearly the major reduction to
$\mathcal{V}(Q_{20})$ comes from the rotational correction
$\epsilon_\mathrm{rot}(Q_{20})$, while the vibrational correction
$\epsilon_\mathrm{vib}(Q_{20})$ produces a smaller, yet not constant,
shift. We also show the inner, $\BI$, outer, $\BII$, barrier and the isomer
excitation energy, $\EII$, defined as illustrated in the figure.

The second ingredient needed for the calculation of the SF lifetimes are the
collective inertias $\mathcal{M}(Q_{20})$. In the present work we evaluated
$\mathcal{M}(Q_{20})$ and $\epsilon_\mathrm{vib}(Q_{20})$ following two
different schemes within the perturbative cranking approximation: the Adiabatic
Time-Dependent HFB theory (ATDHFB)~\cite{Girod1979} and the Gaussian Overlap
Approximation to the Generator Coordinate Method (GOA-GCM)~\cite{Ring1980}:
\begin{align}
    \mathcal{M}_\mathrm{ATDHFB}(Q_{20})&=\frac{\mathrm{M}_{-3}}
    {2(\mathrm{M}_{-1})^2}\,,
    \label{eq:MATD}\\   
    \mathcal{M}_\mathrm{GOAGCM}(Q_{20})&=\frac{(\mathrm{M}_{-2})^2}
    {2(\mathrm{M}_{-1})^3}
    \label{eq:MGCM}\,.
\end{align}
The energy-weighted momentum $\mathrm{M}_{-n}(Q_{20})$ of the quadrupole
generating field can be expressed in terms of the two-quasiparticle excitations
$|\alpha\beta\rangle$:
\begin{equation}
  \mathrm{M}_{-n}(Q_{20})=\sum_{\alpha>\beta}\frac{|\langle\alpha\beta|Q_{20}|
  0\rangle|^2}
  {(E_\alpha+E_\beta)^n}\,,
\end{equation} 
being $E_\alpha+E_\beta$ the excitation energy neglecting the quasiparticle-%
quasiparticle interaction. The ATDHFB scheme has the advantage that it naturally
includes the time-odd response of the system to small perturbations in the
deformation. In the simple case of a center-of-mass motion of the nucleus, the
inclusion of the time-odd momenta allows the ATDHFB scheme to predict the exact
collective inertia as the mass of the nucleus~\cite{Ring1980}. By contrast the
GCM scheme does not include the time-odd response of the system and the
inertias are underestimated unless time-odd momenta coordinates are used as
collective degree of freedom.  However, since translation and fission are
collective phenomena involving different dynamics this argument cannot be used
for claiming a superiority of the ATDHFB scheme over the GCM one.

The $\epsilon_\mathrm{vib}(Q_{20})$ energy is only defined in the GOA-GCM scheme
~\cite{Ring1980}
\begin{equation}
\epsilon_\mathrm{vib}^\mathrm{GOAGCM}(Q_{20})=\frac{G(Q_{20})}
    {\mathcal{M}_\mathrm{GOAGCM}}
    \,.\label{eq:ZPEG}
\end{equation}
However, it is customary \cite{Schunck2016} to also introduce this quantity 
in the ATDHFB scheme by using
the previous expression but replacing the GOA-GCM inertia by the ATDHFB one:
\begin{equation}
    \epsilon_\mathrm{vib}^\mathrm{ATDHFB}(Q_{20})=\frac{G(Q_{20})}
    {\mathcal{M}_\mathrm{ATDHFB}}\,.
    \label{eq:ZPEA} 
\end{equation} 
The quantity $G(Q_{20})$ is the overlap width between two configurations with similar
quadrupole deformations:
\begin{equation}
  G(Q_{20})=\frac{\mathrm{M}_{-2}}{2(\mathrm{M}_{-1})^2}.
\end{equation}

Several calculations of fission cross sections~\cite{Goriely2009} use
collective inertias based on the semiempirical parametrization
$\mu=0.054A^{5/3}$~MeV$^{-1}$ and $\beta_{20}$ as collective variable
of the action integral (\ref{eq:Sl}). This expression reproduces
experimental data in the actinide region for a particular choice of
fission barriers~\cite{Nilsson1968}.  Its validity for heavier nuclei
and/or different barriers is questionable. Hence, we compare 
the spontaneous fission lifetimes obtained with this semiempirical
expression with the results obtained using the ATDHFB and GOA-GCM
approaches. Since the action integral is invariant under uniform
scaling, and $\beta_{20}=\frac{\sqrt{20\pi}}{5A}\frac{Q_{20}}{r^2}$
with $r=1.2A^{1/3}$~fm, we have that:
\begin{equation}
 \mathcal{M}_\mathrm{SEMP}=\mu\left(\frac{d\beta_{20}}{dQ_{20}}\right)^2=
 \frac{0.065}{A^{5/3}}\,\,
 \textrm{MeV}^{-1}\textrm{fm}^{-4}\,\label{eq:MSEM}.
\end{equation}

The upper panels of Fig.~\ref{fig:PES} show the collective inertias obtained
from ATDHFB, GCM and the semiempirical schemes as a function of the quadrupole
deformation. The values are normalized to the inertial mass that corresponds to
the reduced mass of the fragments, $m$. Assuming point like fragments, it can be
expressed as a function of the quadrupole moment as:
\begin{equation}
  \widetilde{\mathcal{M}}(Q_{20})=\frac{m}{\hbar^2}
  \left(\frac{d r}{dQ_{20}}\right)^2 =  \frac{m_N}{4\hbar^{2}Q_{20}}\,,
\end{equation}
where $m_N=938.919$~MeV/$c^{2}$ is the averaged nucleon mass. The fact
that $\mathcal{M}_\mathrm{ATDHFB}$ and $\mathcal{M}_\mathrm{GOAGCM}$ are larger than $\widetilde{\mathcal{M}}$
suggests that the theoretical collective inertias are overestimated in
our calculations, as we will discuss later in Sec.~\ref{sec:bench}.

\begin{figure*}[tb]
 \includegraphics[width=\textwidth]{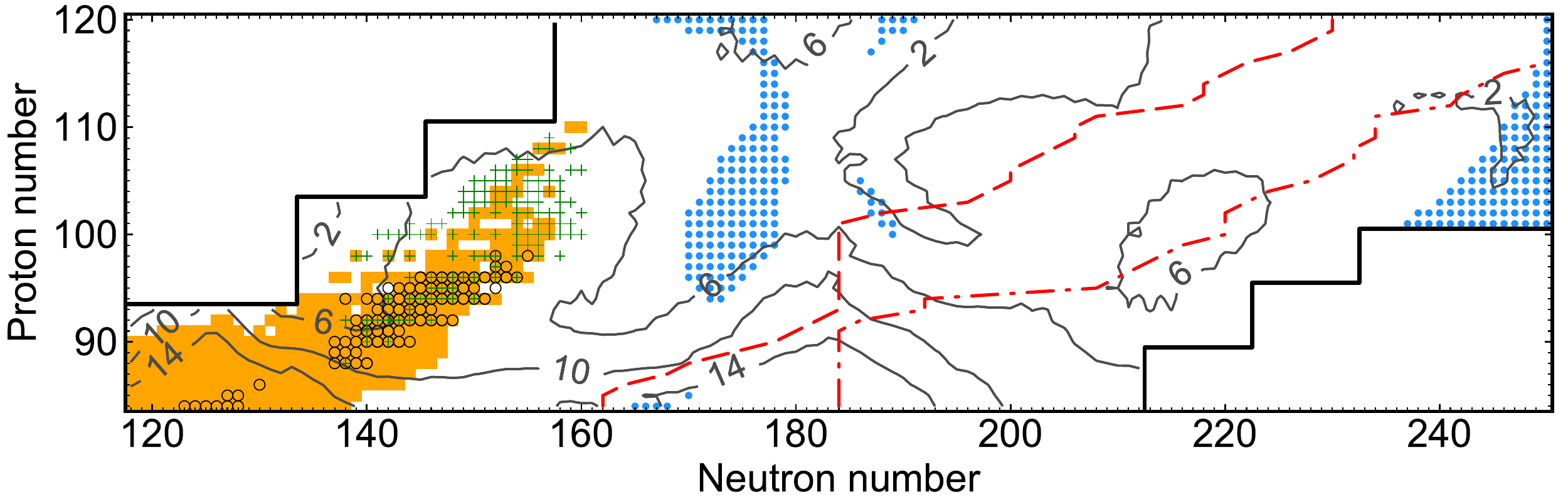}
 \caption{Region of the nuclear landscape explored in this work. 
 Nuclei present in the AME2012 mass table evaluation~\cite{Wang2012} are  
 depicted with orange squares. Nuclei with experimentally measured 
 fission barriers and spontaneous fission lifetimes 
 are marked with open circles and crosses, respectively. Nuclei for which the  BCPM 
 interaction predicts an oblate-deformed ground state are depicted with solid 
 circles. 
 Dashed and dot-dashed lines represent the heaviest  isotope of each element with 
 $S_n\gtrsim2$ and  $S_n\gtrsim0$ MeV,  respectively. Contour lines show 
 the highest predicted fission barrier, in MeV.}
 \label{fig:landscape}
\end{figure*}

Fig.~\ref{fig:landscape} shows the region of the nuclear landscape
explored in this work. Nuclei for which the BCPM interaction predicts
a strong oblate deformation of the ground-state ($\beta_{20}\le-0.1$)
are depicted with solid circles. These nuclei are supposed to undergo
fission through a triaxial path and should be computed with an
explicit breaking of the $K$ quantum number.  Since a triaxial code
for the BCPM interaction is not currently available, we computed the
axial path but it is necessary to keep in mind that predictions of
the fission properties for these nuclei are less reliable.

\subsection{Odd nuclei}\label{sec:odd}
The estimation of nuclear properties of nuclei with an odd number of
protons and/or neutrons is a critical issue in SCMF models. A self-consistent
solution obtained on the same footing as even-even nuclei is rather expensive
from the computational point of view and therefore difficult to be implemented
in systematic calculations (see for example~\cite{Schunck2010,Robledo2012} and
references therein for a general comparison between different approaches). For
this reason, a good compromise in systematic calculations is to use a
phenomenological approach aimed to reproduce the experimental bulk nuclear
properties of odd-even and odd-odd nuclei.

In this work, we computed the bulk nuclear properties of odd nuclei
using the Perturbative Nucleon Addition Method
(PNAM)~\cite{Duguet2001a}.  By adopting the PNAM method BCPM maintains
the same level of accuracy in the calculation of nuclear binding
energies and fission properties for for even-even nuclei and odd-even
and odd-odd nuclei.  Namely, the BCPM EDF without rotational
correction reproduces the experimental binding energies of even-even
nuclei~\cite{Wang2012} with a rms deviation of 2.67~MeV, that
decreases down to 2.37~MeV when odd nuclei computed with the PNAM are
included. But the agreement of the absolute binding energies can be
misleading since the relevant quantities for $r$-process calculations
are the neutron separation energies that determine the
neutron capture cross sections. With the BCPM EDF the neutron
separation energies $S_n$ are reproduced with a rms deviation of
0.30~MeV for even-even nuclei and 0.36~MeV including odd and odd-odd
nuclei. These results gives us the confidence to explore the whole
superheavy landscape using the BCPM EDF in combination with the PNAM.

\begin{figure}[tbh]
  \centering
  \includegraphics[width=0.45\textwidth]{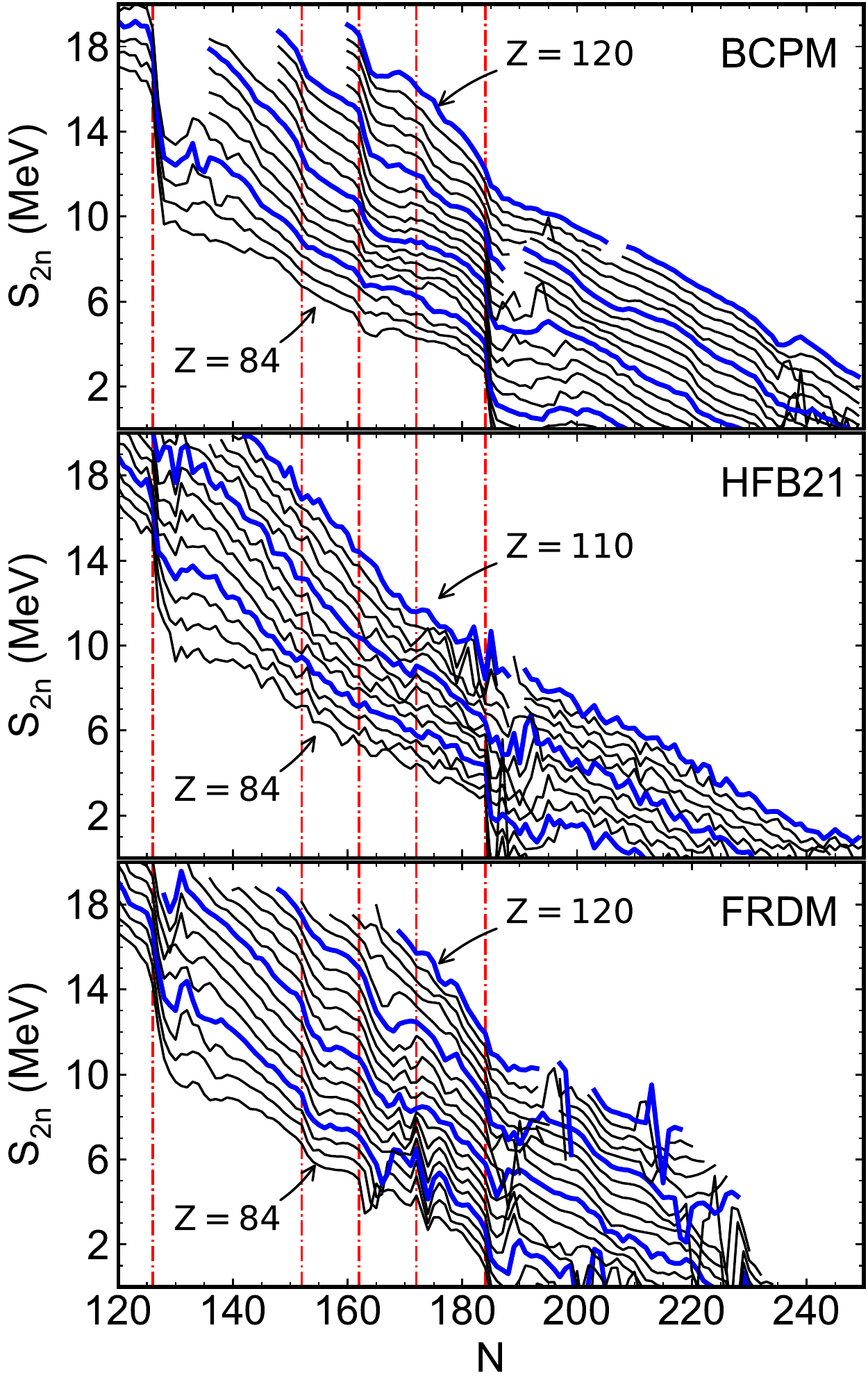}
  \caption{Predicted two-neutron separation energies in MeV for nuclei with
    $84 \le Z \le 120$ as a function of neutron number for three different
    models: BCPM (upper panel), HFB21 (middle panel) and FRDM (lower
    panel).  Isotopic chains are connected by solid lines. Nuclei with a
    proton number equal to 90, 100, 110 and 120 are depicted with blue thick
    lines.
  }
  \label{fig:S2n}
\end{figure}

\begin{figure}[tbh]
  \centering
  \includegraphics[width=0.45\textwidth]{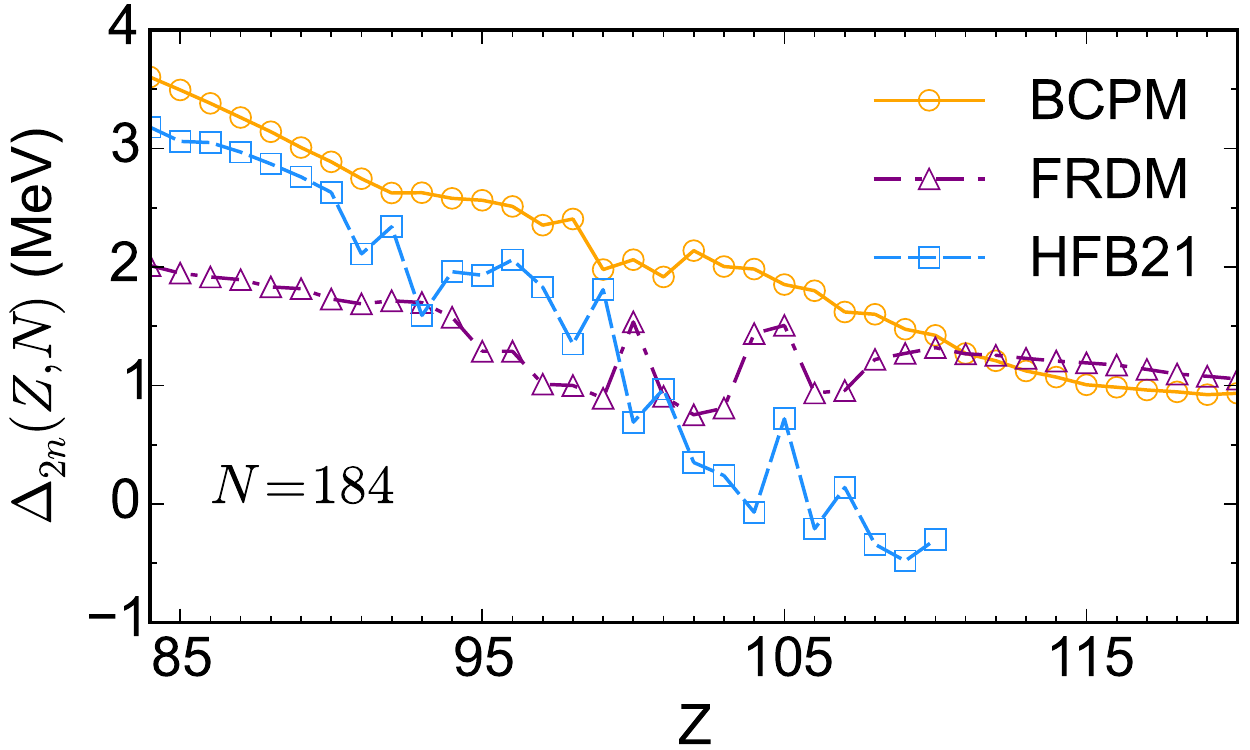}
  \caption{Energy gap $\Delta_{2n}=S_{2n}(Z,N)-S_{2n}(Z,N+2)$ in MeV for
  nuclei with $N=184$ as a function of proton number predicted by BCPM
  (open circles), FRDM (triangles) and HFB21 (squares).
  }
  \label{fig:gap}
\end{figure}

Fig.~\ref{fig:S2n} shows the two-neutron separation energies $S_{2n}$ predicted
by BCPM for nuclei with $84 \le Z \le 120$ and the comparison with
HFB21~\cite{Goriely2010} and FRDM~\cite{Moller1995} models.  Jumps in the
two-neutron separation energies, commonly defined as shell gaps
$\Delta_{2n}(N)=S_{2n}(Z,N)-S_{2n}(Z,N+2)$, are usually associated with the
presence of shell closures and during the $r$-process nucleosynthesis result in
accumulation of matter. For example, for nuclei with $Z \sim 96$ the predicted
neutron magic number $N=184$ plays an important role in the production of the
heaviest elements during the $r$-process nucleosynthesis in neutron star
mergers, since it allows the accumulation of material around
$A\sim280$~\cite{Mendoza-Temis2014}.  Fig.~\ref{fig:gap} shows that BCPM
predicts an higher energy gap compared to HFB17 and FRDM for nuclei with $N=184$
that smoothly disappears with increasing proton number. As we will discuss in
Sec.~\ref{sec:barriers} this disappearance may allow the $r$-path to proceeds
towards heavier nuclei if the fission barriers around $_{100}^{284}$Fm are high
enough.

For some nuclei BCPM predicts a potential energy surface with a vanishing
fission barrier.  For these nuclei a minimum energy cannot be defined and
they are considered unstable, in the sense that after their production they
will immediately decay by fission.  They are followed by nuclei with a
prolate ground state and depicted in Fig.~\ref{fig:S2n} with a gap in the
isotopic lines.

Regarding the collective inertias of odd nuclei, we applied the same
perturbative scheme used
for even-even nuclei described in Sec.~\ref{sec:WKB}. This approach neglects
the enhancement of collective inertias due to the quenching of pairing
correlations in systems with unpaired nucleons, leading to a possible
underestimation of spontaneous fission lifetimes of nuclei with an odd number
of neutrons and/or protons \cite{Rodriguez-Guzman2016}.

\section{Results}\label{sec:results}
\subsection{Benchmarks}\label{sec:bench}
In Ref.~\cite{Giuliani2013} we studied the fission properties of the BCPM EDF for a 
reduced set of even-even nuclei and compared our results with experimental 
measurements. In this paper we present the extension of such calculations to 
the whole superheavy landscape including nuclei with and odd number of 
protons and/or neutrons in the region $84 \le Z \le 120$ and $118 \le  N \le 250$. 

In order to validate the theoretical predictions of the BCPM EDF, we
compared our results of the barriers heights and isomer excitation energies
with the available experimental data of Bj{\o}rnholm and 
Lynn~\cite{Bjørnholm1980} and Capote \emph{et al.}~\cite{Capote2009}.  Whether
the fission barriers can be considered a physical observable or not is
still an argument of discussion in the community. Without entering in this
discussion, we would like to point out that experimental fission barriers
are extracted from fission cross sections measurements, assuming nuclear
level densities and shapes of the potential energy surface predicted by
theoretical models. The experimental values of the fission barriers are
therefore model dependent, and consequently the comparison with theoretical
values should be taken with a grain of salt.

\begin{figure}[tb]
 \includegraphics[width=0.49\textwidth]{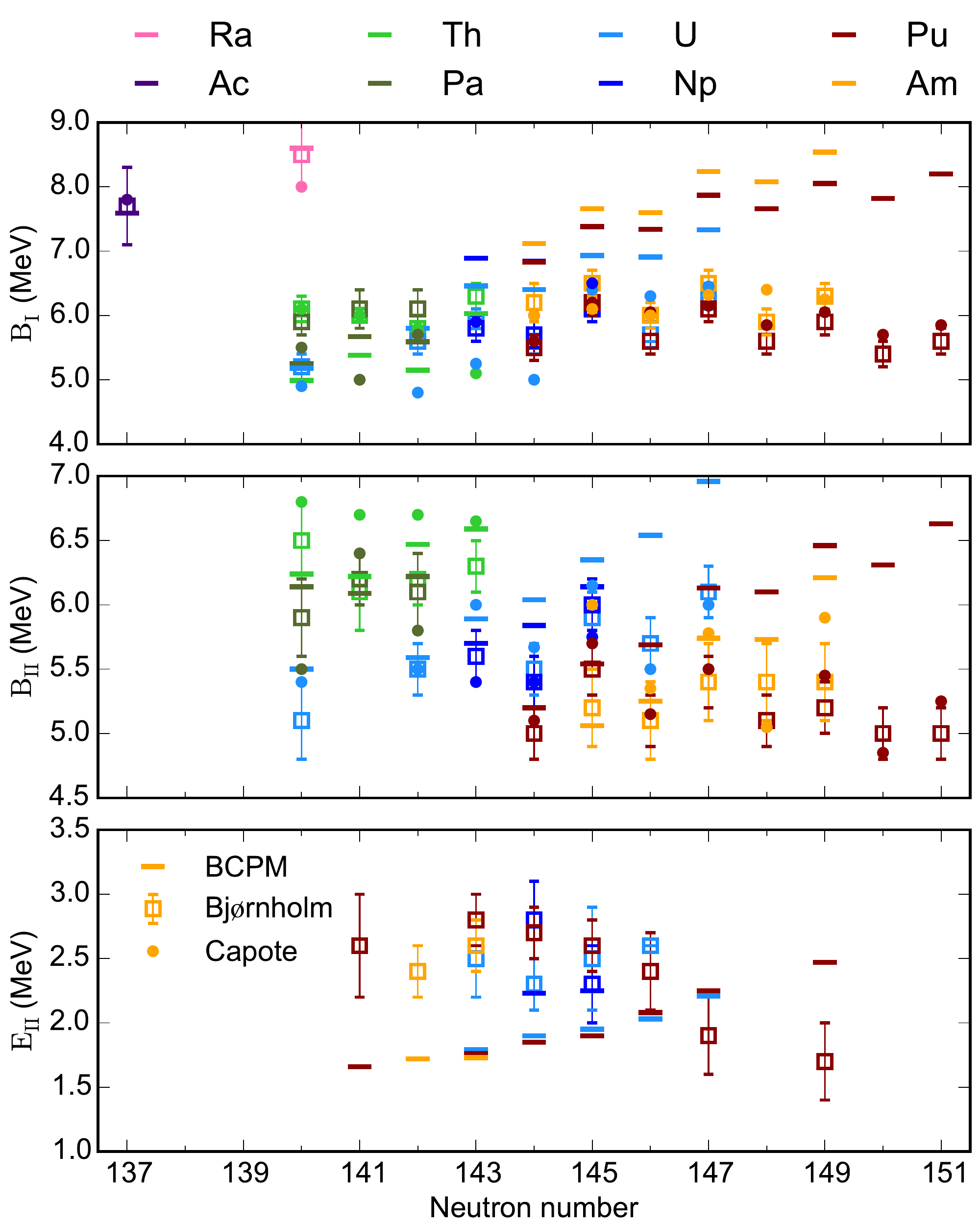}
 \caption{Inner fission barrier height $\BI$  (upper panel), outer fission
   barrier height $\BII$ (middle panel) and isomer excitation energy
   $\EII$ (lower panel) of eight isotopes computed with BCPM (lines) and
   compared with experimental data of Bj{\o}rnholm~\cite{Bjørnholm1980}
   (open squares) and Capote~\cite{Capote2009} (circles).}
 \label{fig:PPES}
\end{figure}

Fig.~\ref{fig:PPES} shows the BCPM predictions and the experimental data of
the inner ($\BI$) and outer ($\BII$) fission barrier height and the isomer
excitation energy ($E_{II}$). We found that BCPM reproduces the $\BI$,
$\BII$ and $E_{II}$ experimental values of Bj{\o}rnholm and
Lynn~\cite{Bjørnholm1980} with a rms deviation of 1.29, 0.81 and 1.22~MeV
respectively. The discrepancies with the data set of Capote \emph{et
al.}~\cite{Capote2009} are slightly larger: 1.51~MeV for $\BI$ and 0.97~MeV
for $\BII$, while no data is available for $\EII$. The largest differences
have been found in the uranium, plutonium and americium isotopes. For
these nuclei BCPM predicts an increase of the three quantities with
increasing neutron number while experimental data shows an almost constant
trend. Due to the imposition of axial symmetry
the inner fission barrier heights are overestimated up to 2.5~MeV, which is in 
agreement with recent works studying the impact of triaxiality in fission
calculations~\cite{Erler2012a,Staszczak2013,Rodriguez-Guzman2014}.
In many cases the PNAM provides a good description of the odd-even
staggering of the fission barriers within single isotopic chains, and the
inclusion of odd nuclei does not substantially modify the agreement with
experimental data.

The other main spontaneous fission observable besides the barrier
height is the spontaneous fission lifetime $\tsf$, that can be
experimentally measured without any model assumption. This observable
can be estimated with logarithmic precision within the semiclassical
WKB formalism described in Sec.~\ref{sec:WKB} and is closely related
to the penetration probability, an important ingredient in the
evaluation of fission cross sections.

\begin{figure*}
 \includegraphics[width=\textwidth]{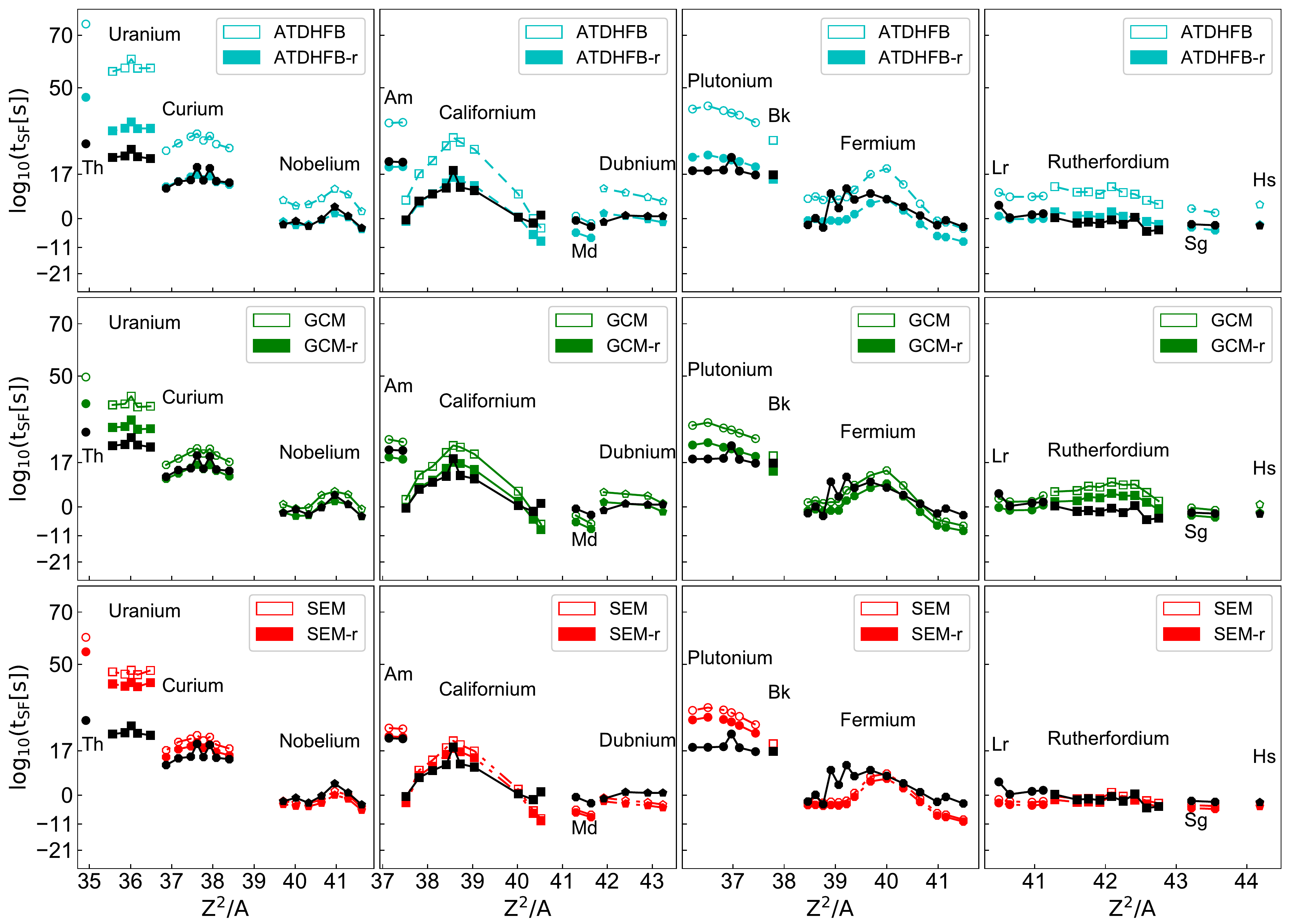}
 \caption{
   {Spontaneous fission lifetimes $\tsf$ as a function of the
   fissibility parameter. Experimental data (black solid symbols) is compared
   with different inertia schemes: ATDHFB (upper panels), GCM (middle panels) and
   semiempirical formula (lower panels). The results obtained with the
   renormalized collective inertia (see Table~\ref{tab:tsf}) are depicted with
   colored full symbols.}}
   \label{fig:tsf-exp}
\end{figure*}

We computed the $\tsf$ using the three different collective inertia schemes:
ATDHFB (Eq.~(\ref{eq:MATD})), GCM (Eq.~(\ref{eq:MGCM})) and the semiempirical
formula (Eq.~(\ref{eq:MSEM})).  Fig.~\ref{fig:tsf-exp} shows the comparison
between our results and the experimental data extracted from
Ref.~\cite{Holden2000,Khuyagbaatar2008}. Both the ATDHFB and GCM microscopic
schemes predict larger SF lifetimes compared to the experimental data, while the
semiempirical formula overestimates the $\tsf$ of nuclei with $Z\le98$ and
slightly underestimates the lifetimes of nuclei with $Z\ge99$. Due to the
inclusion of the time-odd response of the nucleus to small perturbations of the
deformation, the $\tsf$ predicted by the ATDHFB scheme are systematically larger
than the GCM ones. 

Fig.~\ref{fig:tsf-exp} shows that the spread among theoretical lifetimes and the
discrepancy with experimental data are large for light actinides, while for
heavier nuclei predictions become more accurate and precise.  This convergence
of theoretical calculations can be understood by looking at the fission barriers
plotted in Fig.~\ref{fig:PES}. The left panel shows the fission path of the
nucleus $^{232}$Th giving the largest difference between theoretical and
experimental half-lives.  This nucleus presents a broad fission barrier together
with a large collective inertia, resulting in a large action integral $S(L)$
where variations in the collective inertias have a strong impact in the
spontaneous fission lifetimes. On the other hand, the nucleus $^{262}$No has a
much shorter barrier with relatively small inertia between the classical turning
points. This configuration reduces the value of the action integral and the
impact of different collective inertias schemes in the absolute  magnitude of
$\tsf$, all giving a good prediction of the experimental value.  For $r$-process
nuclei where fission may play a relevant role, like $^{290}$No and $^{316}$Ds
plotted in Fig.~\ref{fig:PES}, the $\tsf$ has to be relatively short.  Therefore
these nuclei must have a narrow and/or low fission barrier and small collective
inertias, bringing to a level of precision in the estimation of the $\tsf$
closer to the one obtained for the $^{262}$No rather than the $^{232}$Th.

For most of the isotopic chains the general trend of the spontaneous
fission lifetimes is well reproduced by all the collective inertias
schemes. Moreover, the odd-even staggering of the fission barriers is
reflected in the lifetimes in a rather good agreement with experimental
data. However it is important to notice that this staggering is more
pronounced in the experimental $\tsf$, suggesting for a missing mechanism
enhancing the collective inertias in these nuclei as discussed in Sec.
\ref{sec:odd}. 

The general overestimation of the spontaneous fission lifetimes in 
Fig.~\ref{fig:tsf-exp} suggests that both the fission barriers and the
collective inertias are overestimated in our approach. We already traced back the
overestimation of the fission barriers to the imposition of axial symmetry 
and in the following we will discuss the origin of the  
overestimation of the collective inertias in our calculations.

In this work the fission path was obtained as a function of the quadrupole
moment by minimizing the potential energy (see Fig.~\ref{fig:PES}). The
simplicity of this scheme allows for a systematic calculation of fission
properties through the whole superheavy landscape. However, it does not
incorporate important dynamic effects that appear when the fission path is
obtained by minimizing the collective action. As it was studied in several
recent papers~\cite{Giuliani2014,Sadhukhan2014,Zhao2016a} the dynamic approach
strongly reduces the collective inertia and the $\tsf$, improving the agreement
with experimental data in actinides. Unfortunately such kind of studies
exploring multidimensional energy surfaces with several degrees of freedom are
extremely expensive from a computational point of view and cannot be applied to
systematic calculations like the one presented in this paper.
Therefore we propose a renormalization of the collective
inertias aimed to take into account the dynamics neglected in
the static one-dimensional picture and improve the agreement with
experimental data. The normalization coefficient is obtained by minimizing the
deviation between theoretical calculations and experimental data of the
spontaneous fission lifetimes for each scheme presented in Sec.~\ref{sec:WKB}. 
Since the $\tsf$ can vary in many orders of
magnitude, we follow the prescription of Ref.~\cite{Bertsch2015b} and
use the logarithm of the ratio of theory to experiment
\begin{equation}
  R_\tau=\log\left(\frac{t_\mathrm{sf}^\mathrm{BCPM}}{t_\mathrm{sf}^\mathrm{exp}}\right)\,.
\end{equation}
The target performance $\bar{R}_\tau$ and the variance  $\sigma_\tau$ are then
obtained as:
\begin{eqnarray}
  \bar{R}_\tau&=&\frac{1}{N}\sum_{i=1}^N R_{\tau,i}\,,\\
  \sigma_\tau&=&\frac{1}{N}\left(\sum_{i=1}^N (R_{\tau,i}-\bar{R}_\tau)^2\right)^{1/2}\,,
\end{eqnarray}
being $N$ the number of nuclei used in the benchmark.  Comparing the logarithm
of the ratio of theory to experiment  we found that the target performance of
the ATDHFB, GCM and semiempirical inertia schemes is 11.583, 4.691 and 2.036,
respectively.  We find that the minimum value of the target performance is
obtained by multiplying the ATDHFB, GCM and SEMP collective inertias by a factor
0.497, 0.731 and 0.868 respectively (labeled as ATDHFB-r, GCM-r and SEMP-r in
Fig.~\ref{fig:tsf-exp} and Table~\ref{tab:tsf}).  Fig.~\ref{fig:tsf-exp} shows
that this renormalization strongly reduces the $\tsf$ of light actinides (Th, U
and Pu), where the high stability against the fission process leads to large
values of the action integral and larger discrepancies with the experimental
data. On the other hand, as we move towards heavier nuclei the impact of the
renormalization in $\tsf$ tend to decrease.  This is shown in
Fig.~\ref{fig:tsf-exp} where the differences between non-renormalized and
renormalized $\tsf$ are smaller for heavier nuclei as compared to lighter ones.
The renormalization of the collective inertias, which is 
justified by the
inclusion of dynamic effects beyond static calculations, shall be considered as
an alternative to other approaches like the renormalization of the fission
barriers proposed in Ref.~\cite{Goriely2009} aimed to improve the agreement
between theoretical calculations and experimental data.
Dynamic effects also tend to increase the rotational and vibrational
corrections of Eq.~\eqref{eq:Veff} but to a much lesser extent than the effect on
the inertias. This is the reason why only the inertias have been renormalized.

\begin{table}[tb]
\begin{tabular}{crr}
\toprule
$\mathcal{M}$   & $\bar{R}_\tau$  & $\sigma_\tau$ \\
\hline
ATDHFB          & 11.583            & 6.447         \\
GCM             &  4.691            & 4.236         \\
SEMP            &  2.036            & 6.126         \\
\hline
ATDHFB-r        & -0.007            & 3.403         \\
GCM-r           & -0.006            & 3.339         \\
SEMP-r          &  0.004            & 5.231         \\
\toprule
\end{tabular} 
\caption{Target performances ($\bar{R}_\tau$) and variances ($\sigma_\tau$) of the 
spontaneous fission lifetimes obtained with ATDHFB, GCM and semiempirical collective 
inertias described in Sec.~\ref{sec:WKB}. The lower table shows the results obtained 
by multiplying the collective inertias by a renormalization factor (0.497 for ATDHFB, 
0.731 for GCM and 0.868 for SEMP). Experimental values extracted from 
Ref.~\cite{Holden2000,Khuyagbaatar2008}. 
\label{tab:tsf}}
\end{table}

The results of the fission barrier height and the SF lifetimes presented in
this section proved the capability of the BCPM+PNAM scheme to
reproduce the experimental data. We consider this agreement rather
satisfactory, specially taking into account that the BCPM EDF was
fitted in order to reproduce the nuclear masses of the AME2003 mass
table~\cite{Audi2003} and it does not contain any information
regarding the fission properties of superheavy nuclei.

\subsection{Systematic of fission barriers}\label{sec:barriers} 
\begin{figure}[tb]
 \includegraphics[width=0.49\textwidth]{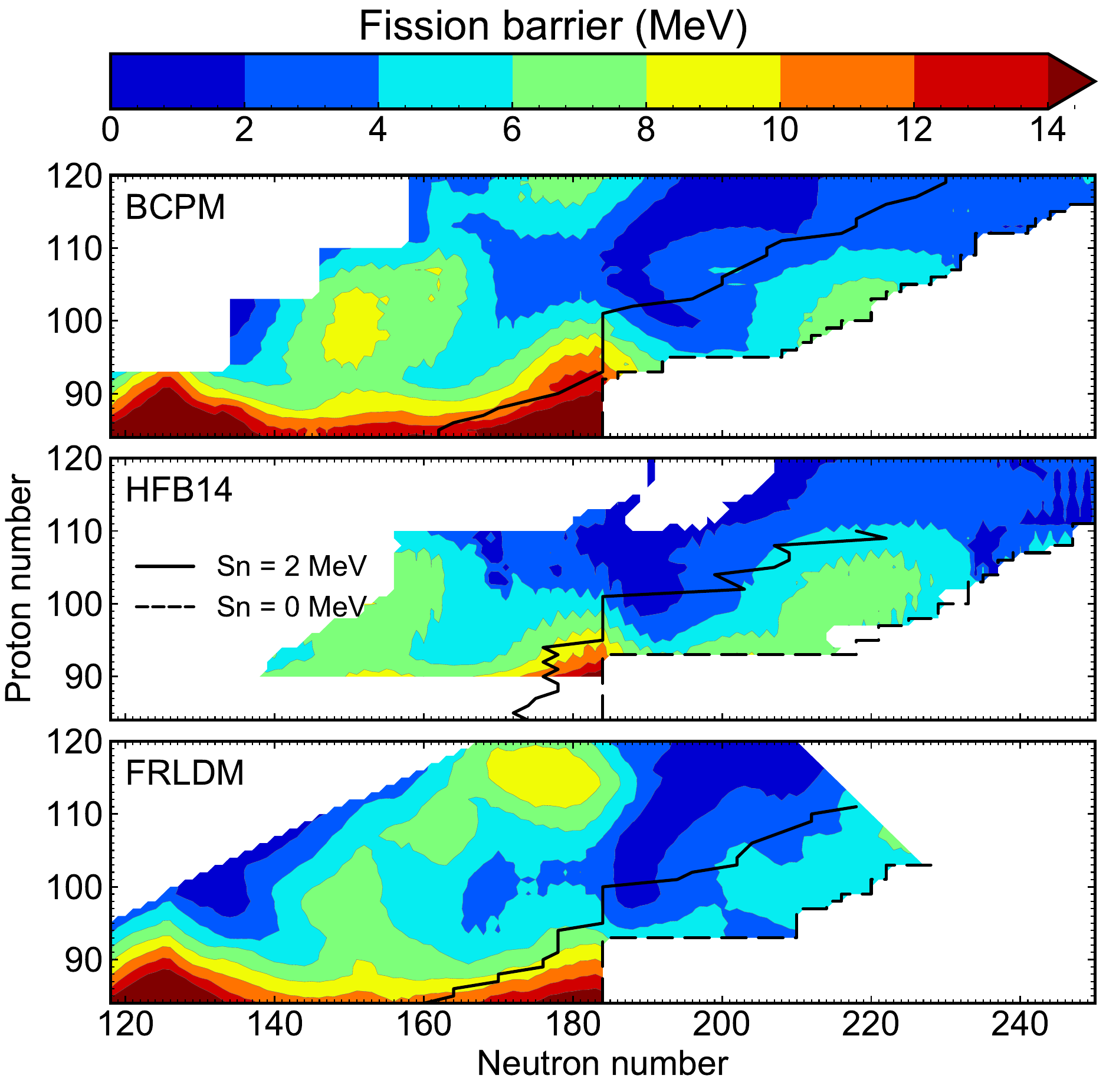}
 \caption{Calculated fission barrier heights in MeV in the region 
  $84 \le Z\le 120$ 
  and $118  \le  N \le 250$ for three different mass models: 
  BCPM (this study, upper panel), HFB14~\cite{Goriely2009} (middle panel) 
  and FRLDM~\cite{Moller2015} (lower panel). Drip lines are represented 
  by dashed black lines. The solid black lines show 
  the $r$-process path, given by the heaviest isotope of each nuclei with 
  $S_n\ge2\,$MeV.}
 \label{fig:Bf-comp}
\end{figure}

Fig.
\ref{fig:Bf-comp} shows the highest barrier predicted by the BCPM EDF for
nuclei in the region  $84 \le Z \le 120$ and $118 \le  N \le 250$ and the
comparison with the theoretical predictions of the FRLDM~\cite{Moller2015}
and HFB-14~\cite{Goriely2009} nuclear models.

The general trend of fission barriers gives a crude estimation of the stability
of nuclei against the fission process and reflects the impact of shell closures.
BCPM predicts five different islands of local maximum placed around nuclei
$^{210}_{84}$Po, $^{268}_{84}$Po, $^{250}_{100}$Fm, $^{320}_{102}$No and
$^{300}_{120}$Ubn. The increase of fission barriers around  $^{210}_{84}$Po and
$^{268}_{84}$Po is related with the presence of the neutron magic numbers $
N=126$ and $ N=184$, leading to spherical nuclei with fission barriers up to
24-25 MeV. These two islands are separated by prolate nuclei with fission
barriers around 12 MeV and a group of slightly oblate nuclei around $ Z/N
=86/176$. Regions around $^{250}_{100}$Fm and $^{300}_{120}$Ubn are usually
referred to as the ``peninsula of known nuclei'' and ``island of stability'',
respectively~\cite{Chowdhury2008}.  The peninsula is formed by prolate-deformed
nuclei with fission barriers between 6 and 9 MeV and it extends up to $ Z/N
\approx110/166$. On the other hand, nuclei in the island of stability are either
oblate (for lower $ N$) or spherical (higher $ N$) with fission barriers around
7 MeV. The peninsula and the island are separated by a rather narrow region of
prolate nuclei with $ A \sim280$ where the fission barriers decrease to 3-5 MeV.
Finally, the region around $^{320}_{102}$No is formed by strongly-deformed
nuclei ($\beta_{20}\sim0.25$) with barriers between 7 and 8 MeV. BCPM predicts a
region of vanishing barriers around $ Z/N =116/208$ and for nuclei with $ A
\sim292$ around $ N\sim188$. As discussed later, this region of vanishing
barriers may play an important role for terminating the $r$ process via neutron
induced fission.

For a complete comparison Fig.~\ref{fig:Bf-comp} shows the fission
barriers predicted by two other models: the macroscopic-microscopic
Finite Range Liquid Drop Model (FRLDM)~\cite{Moller2015} and the
self-consistent mean field approach based on the HFB14 Skyrme
parametrization~\cite{Goriely2009}.  Both models show a general trend
of the fission barriers similar to the one obtained with the BCPM EDF,
with two islands of larger fission barriers around $ Z/N =90/184$ and
$ Z/N =100/150$. Moreover, all the models predict a region of
vanishing barriers around $ Z/N =104/188$. BCPM tends to predict
larger barriers compared to those obtained with FRLDM for nuclei with
$Z\le100$ corresponding to the neutron magic number $N=184$. On the
other hand, FRLDM predicts larger barriers (up to 5 MeV) in the region
around $Z/N =112/178$.  Comparing the results obtained with BCPM and
HFB14, we found these two mean-field models predict similar maximum
fission barrier heights. The rms deviation between both models is
1.03~MeV, while the deviation between BCPM and FRLDM is 2.31~MeV. The major
differences between BCPM and HFB14 are found in neutron rich actinides where
HFB14 predicts fission barriers larger by 2-3~MeV, and around $Z/N =97/187$ and
$106/196$ where BCPM barriers are roughly 2~MeV larger. It is clear then, that
different models predict quantitatively different fission barriers in the region
far from stability where the $r$ process occurs. However, it is not possible to
determine a priori which model is the preferable one. Therefore, calculations of
fission reaction rates obtained from different nuclear models are required to
assess the sensitivity of the $r$-process abundances to uncertainties in the
estimation of the fission properties of superheavy nuclei.

\begin{figure}[tb]
 \includegraphics[width=0.49\textwidth]{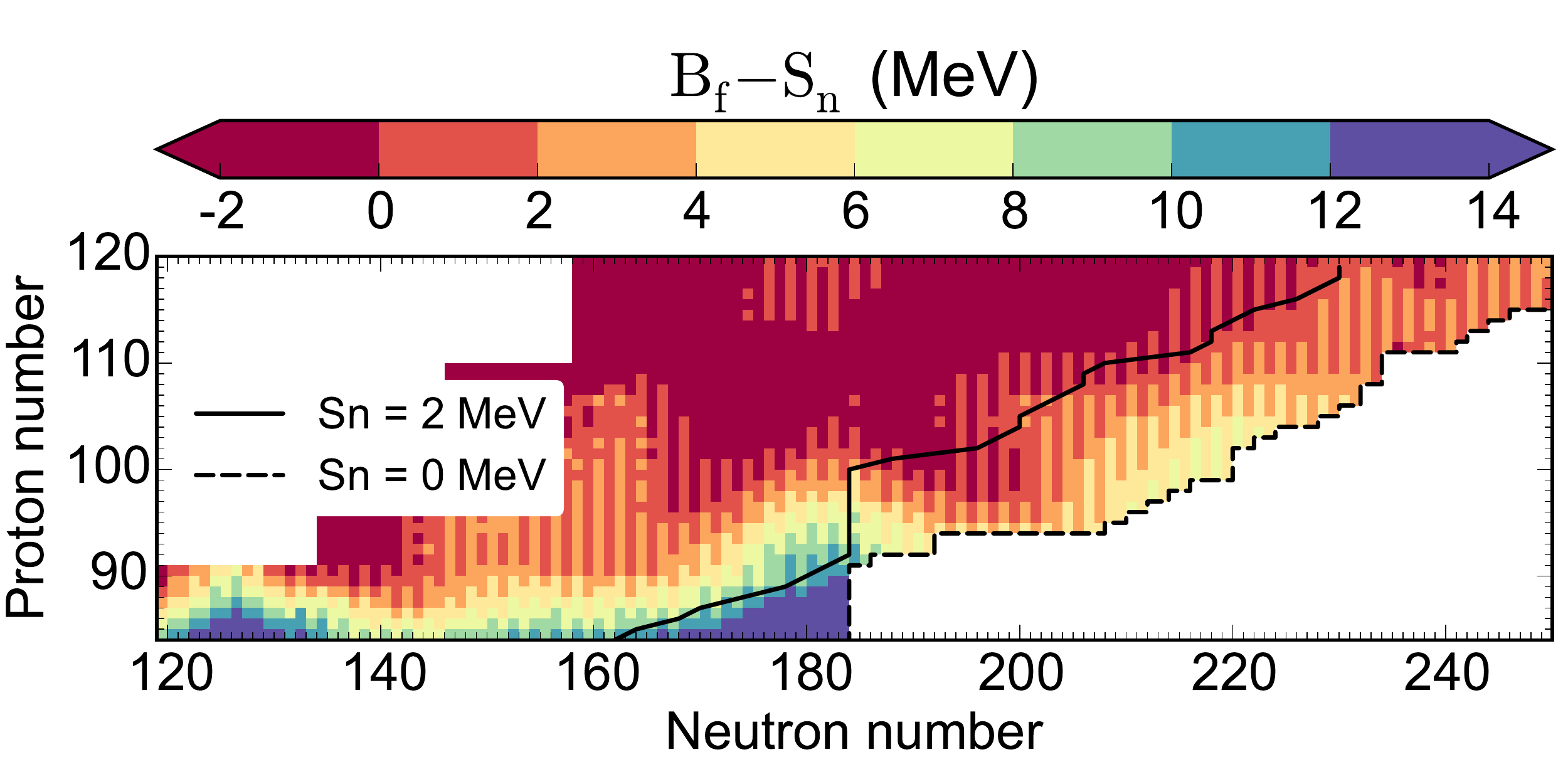}
 \caption{Energy window for the neutron-induced fission $B_f-S_n$ 
 computed with the BCPM EDF.  The solid  black line represents 
 the $r$-process path, given by the heaviest isotope of each nuclei 
 with  $S_n\ge2\,$MeV. The drip line predicted by the BCPM EDF 
 is represent by dashed line.}
 \label{fig:Bf-Sn}
\end{figure}

Another quantity of major interest for astrophysical calculations is the energy
window for neutron-induced fission given as the difference between the highest
fission barrier height and the neutron separation energy $B_f-S_n$. This
quantity indicates whether the production of superheavy nuclei during the $r$
process can be inhibited by neutron-induced fission, recycling the material to
lighter fission products.  Fig.~\ref{fig:Bf-Sn} shows the values of $B_f-S_n$
obtained with the BCPM EDF. In principle, an appropriate estimation of the
$r$-process path would require a network calculation taking into account neutron
captures, beta decays and photodissociations. However, from simple arguments it
is still possible to make a rough estimation of where the $r$-process path will
be terminated by the neutron-induced fission. For typical astrophysical
conditions in neutron star mergers, the $r$-process path is supposed to proceed
along nuclei with constant neutron separation energy $S_n\sim
2$--3~MeV~\cite{Martinez-Pinedo2008}.  On the other hand, the excitation energy
of a nucleus after capturing a neutron is given by the neutron separation
energy.  From these arguments one concludes that nuclei with $B_f-S_n\sim2\,$MeV
will immediately fission after capturing a
neutron~\cite{Panov2009a,Petermann2012}.  Fig.  \ref{fig:Bf-Sn} shows how the
$r$-process path is stacked along nuclei with $ N=184$ until $Z=102$, where the
disappearance of the jump in the neutron separation energy described in
Sec.~\ref{sec:odd} allows to overcome the waiting point.  However, at this point
the $r$-process path has already proceeded into the region of low fission
barriers where $B_f-S_n$ drops below zero ($Z/N =102/190$). Therefore, we may
expect the $r$-process nucleosynthesis of superheavy nuclei to be terminated by
the neutron induced fission in the region around $ Z/N =102/190$.

\subsection{Systematic of fission lifetimes}\label{sec:tsf}

The trend of the fission barriers gives only a rough hint of the stability
of the nucleus against the fission process. As it was already explained in
Sec.~\ref{sec:WKB}, the probability of the system to penetrate the fission
barrier is determined by a complex process where several ingredients must
be taken into account and it can not be solely determined by the height of
the barrier. A more complete picture can be therefore obtained studying the
trend of the spontaneous fission lifetimes and the contribution of the
different terms entering in Eq.~(\ref{eq:Sl}). In this section we will
study the sensitivity of the spontaneous fission lifetimes to variations in
the collective inertias $\mathcal{M}(Q_{20})$, the vibrational energy
corrections $\epsilon_\mathrm{vib}(Q_{20})$ and the collective ground state
energy $E_0$.

\begin{figure}[tb]
 \includegraphics[width=0.49\textwidth]{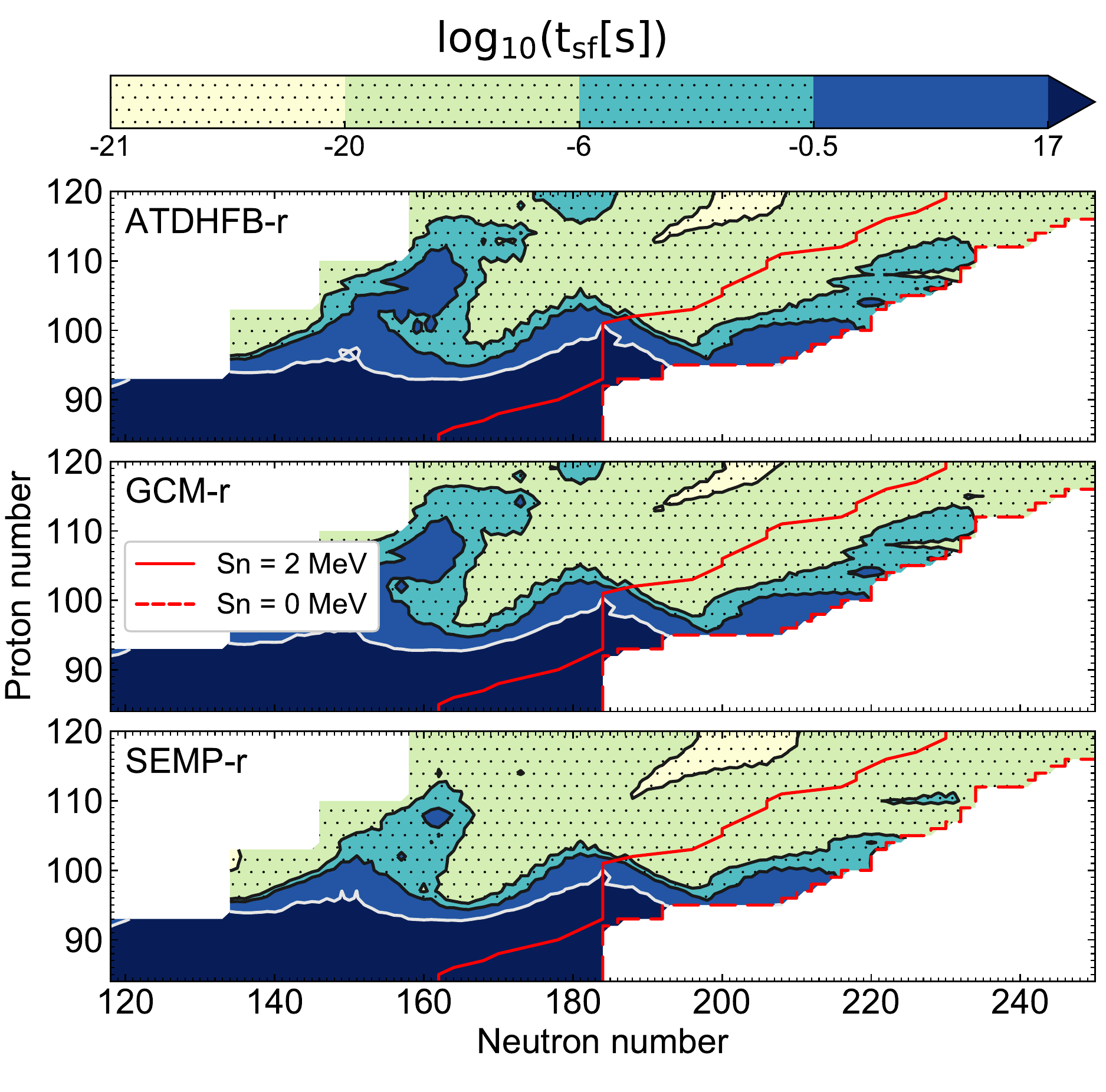}
 \caption{Decimal logarithm of the spontaneous fission 
   lifetimes computed with the ATDHFB-r (upper panel), GCM-r (middle panel) and
   SEMP-r (lower panel) schemes (see Table~\ref{tab:tsf} for the renormalization
   coefficients). Dotted regions represent nuclei with $\tsf\le3\,$s which may
   be relevant for $r$-process calculations. All the lifetimes were obtained
   with $E_0=0.5$ MeV.}
 \label{fig:tsf-all}
\end{figure}

Fig.~\ref{fig:tsf-all} shows the $\tsf$ obtained from Eq.~(\ref{eq:Sl}) using
the three different schemes of the collective inertias described in
Sec.~\ref{sec:WKB} and renormalized using the coefficients of
Sec.~\ref{sec:bench}.  For the ATDHFB and GCM schemes the vibrational energy
corrections are consistently computed using Eq.~(\ref{eq:ZPEA})
and~(\ref{eq:ZPEG}). For the semiempirical inertias we arbitrarily choose the
$\epsilon_\mathrm{vib}^\mathrm{GCM}(Q_{20})$ scheme.  Regarding the collective
ground state energy, all the lifetimes were obtained with $E_0=0.5$~MeV. Due to
the arbitrariness in the choice of these last two parameters, the second part of
this section will be devoted to study the sensitivity of the lifetimes on
$\epsilon_\mathrm{vib}$ and $E_0$.

From Fig.~\ref{fig:tsf-all} it is possible to conclude that the trend
of the spontaneous fission lifetimes resembles the general trend of
the maximum fission barrier height plotted in the upper panel of 
Fig.~\ref{fig:Bf-comp}. This means that quantities like collective inertias
and the shape of the barrier are responsible for local variations in
the stability of the nucleus against the fission
process. The three schemes predict a region around $Z/N=120/200$ with lifetimes
of the order $\tsf \sim 10^{-21}\,$s, corresponding to nuclei with almost
vanishing fission barriers.

\begin{figure}[tb]
 \includegraphics[width=0.49\textwidth]{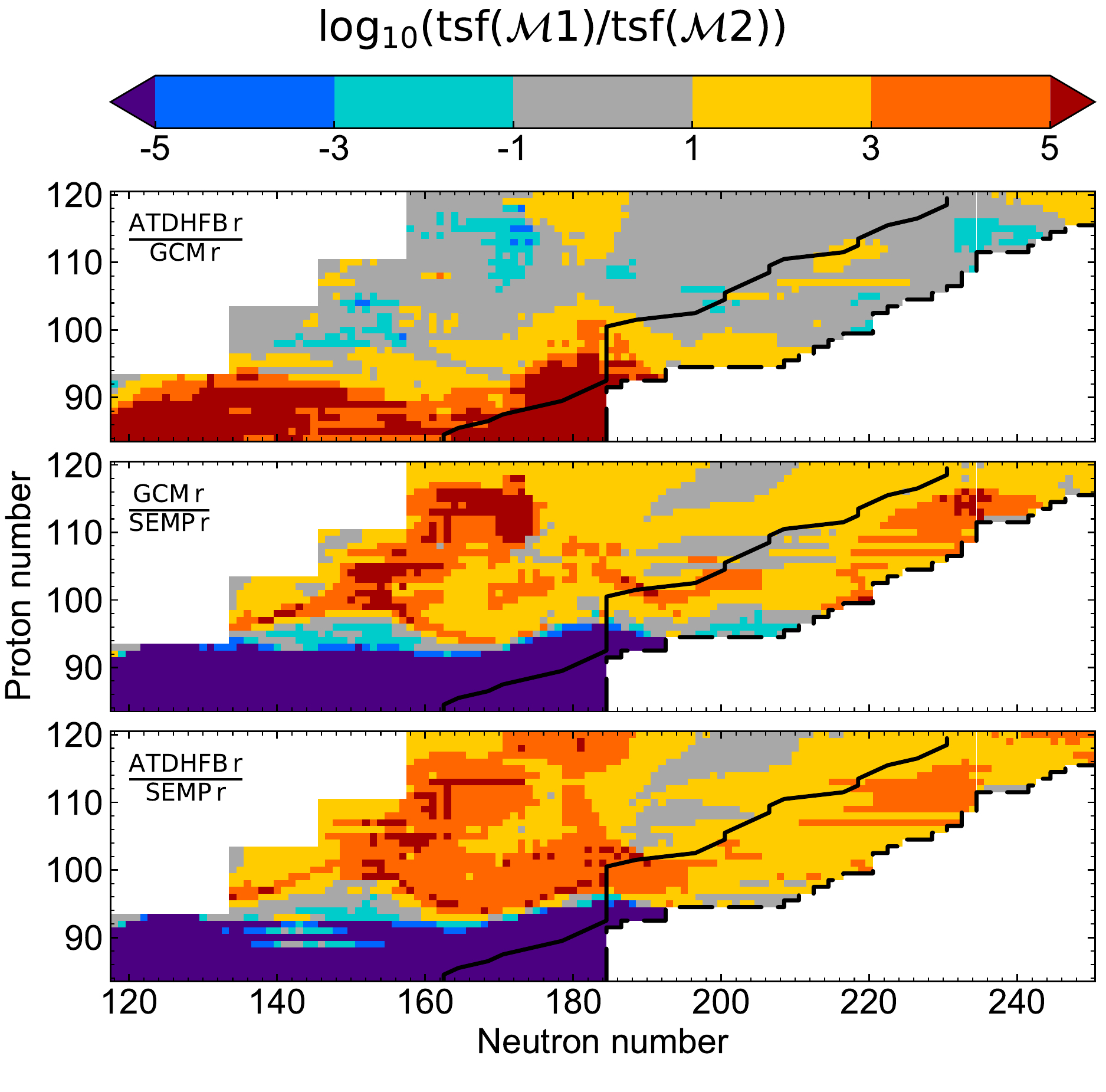}
 \caption{Logarithm of the ratio of the spontaneous fission lifetimes 
 for different combinations of 
 collective inertias: $R^\mathrm{ATDHFB}_\mathrm{GCM}$ (upper panel);
 $R^\mathrm{GCM}_\mathrm{SEMP}$
 (middle panel) and $R^\mathrm{ATDHFB}_\mathrm{SEMP}$ (lower panel).}
 \label{fig:tsfM}
\end{figure}

For a more detailed comparison between the different models Fig.~\ref{fig:tsfM}
shows the ratio of the lifetimes computed with different collective inertia
schemes: ATDHFB to GCM (upper panel), ATDHFB to semiempirical inertias (middle
panel) and GCM to semiempirical inertias (lower panel). The values showed in
this plot correspond to the quantity
\begin{equation}
  R^{\mathcal{M}1}_{\mathcal{M}2}=\log\left(\frac{t_\mathrm{sf}(\mathcal{M}_1)}
  {t_\mathrm{sf}(\mathcal{M}_2)}\right)\,,
\end{equation}
being $t_\mathrm{sf}(\mathcal{M}_i)$ the spontaneous fission lifetime
computed using the collective inertia $\mathcal{M}_i$.
Fig.~\ref{fig:tsfM} shows that the largest variations among the different
schemes are found in nuclei with $Z\lesssim96$, where the spontaneous fission
lifetimes are large. This is because in these nuclei differences in the
collective inertia strongly impact the absolute value of the action integral
$S(L)$ entering in the exponential of the lifetimes $\tsf$.  In the $r$ process
we are interested in nuclei with $\tsf\le3$~s (marked as the dotted region in
Fig.~\ref{fig:tsf-all}), since this is the average timescale at which the $r$
process operates from the onset of neutron captures till the exhaustion of all
neutrons. Fig.~\ref{fig:tsfM} shows that for these nuclei the variations in the
$\tsf$ predicted by the different collective inertia schemes are usually below 3
orders of magnitude, much lower than for the rest of the nuclei. By
renormalizing, the inertias we were able to reduce the sensitivity of the
spontaneous fission lifetimes, specially in the case of the ATDHFB and GCM
schemes. In the latter cases the difference for most of the $r$ process nuclei
is less than one order of magnitude.

\begin{figure}[tb]
 \includegraphics[width=0.49\textwidth]{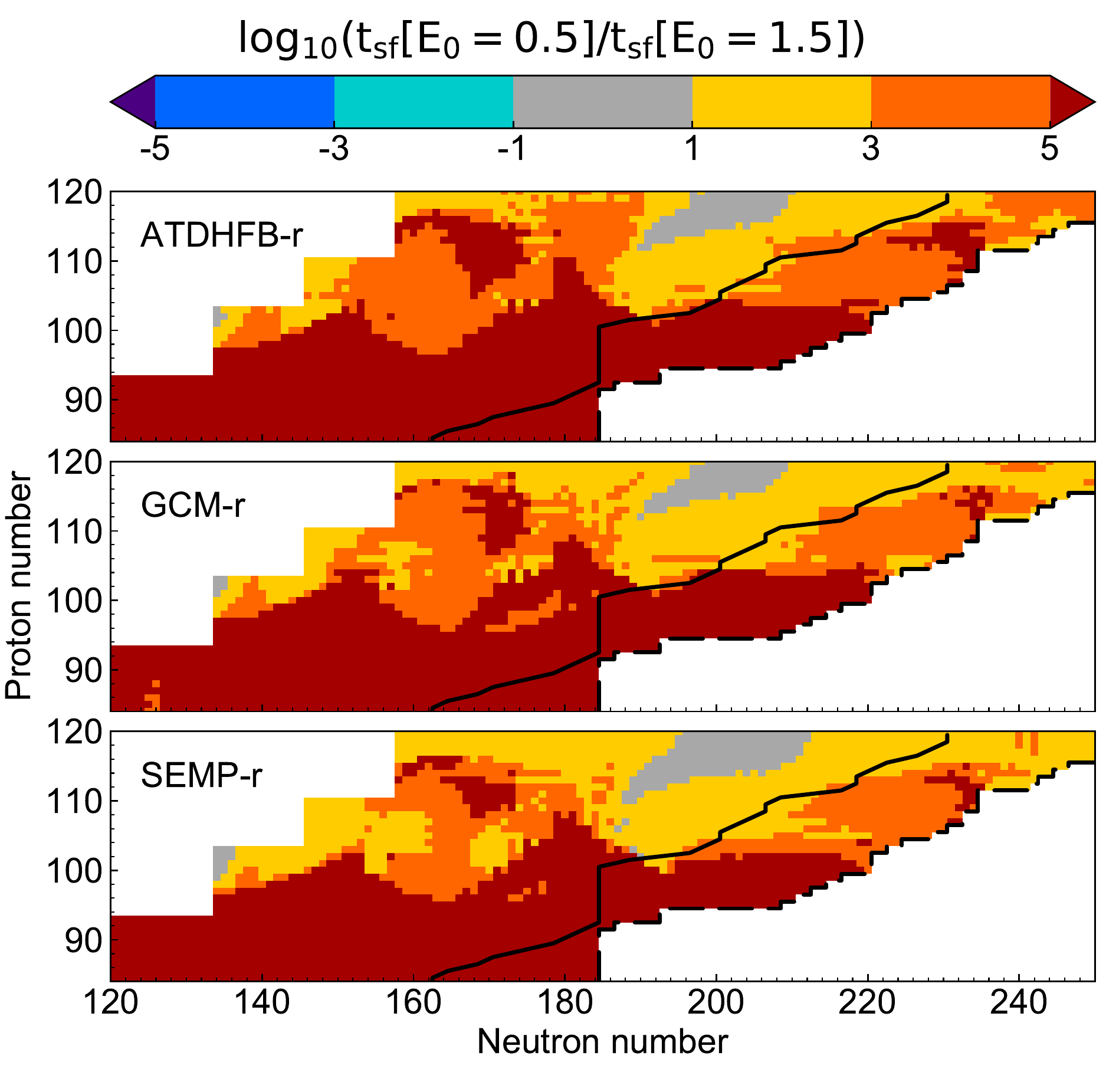}
 \caption{Sensitivity of the spontaneous fission lifetimes to 
 different values of the collective ground-state energy 
 $\log_{10}\left[{t_\mathrm{sf}(E_0=0.5\,
 \mathrm{MeV})}/{t_\mathrm{sf}(E_0=1.5\,\mathrm{MeV})}\right]\,$ 
 computed with different collective inertias:  ATDHFB (upper panels), 
 GCM (middle panels) and 
 semiempirical inertia formula (lower panels).}
 \label{fig:tsfE0}
\end{figure}

To better quantify the robustness of the spontaneous fission lifetimes
Fig.~\ref{fig:tsfE0} shows the sensitivity of $\tsf$ to variations in the
collective ground state energy. We notice that by increasing $E_0$ by 1.0~MeV,
the lifetimes can vary by more than 5 orders of magnitude even in nuclei with
relatively short lifetimes and close to the $r$ process path. The reason is that
nuclei with low barriers can still have a complex shape presenting multiple
humps, like for instance the case of the $^{290}$No plotted in
Fig.~\ref{fig:PES}. As it was already extensively studied in
Ref.~\cite{Warda2002}, the presence of a second fission isomer increases the
spontaneous fission lifetimes by several orders of magnitude. By increasing the
$E_0$ to 1.5~MeV the isomer can be shifted below the ground-state energy of the
nucleus, and therefore the lifetimes are strongly reduced since the third hump
does not contribute anymore to the penetration probability.  Assuming that
changes to $E_0$ have a similar impact in the action integral as the
renormalization of the fission barrier, the latter may affect the location of
the fissioning region for the $r$ process nucleosynthesis more than the
renormalization of the collective inertias.

\begin{figure}[tb]
 \includegraphics[width=0.49\textwidth]{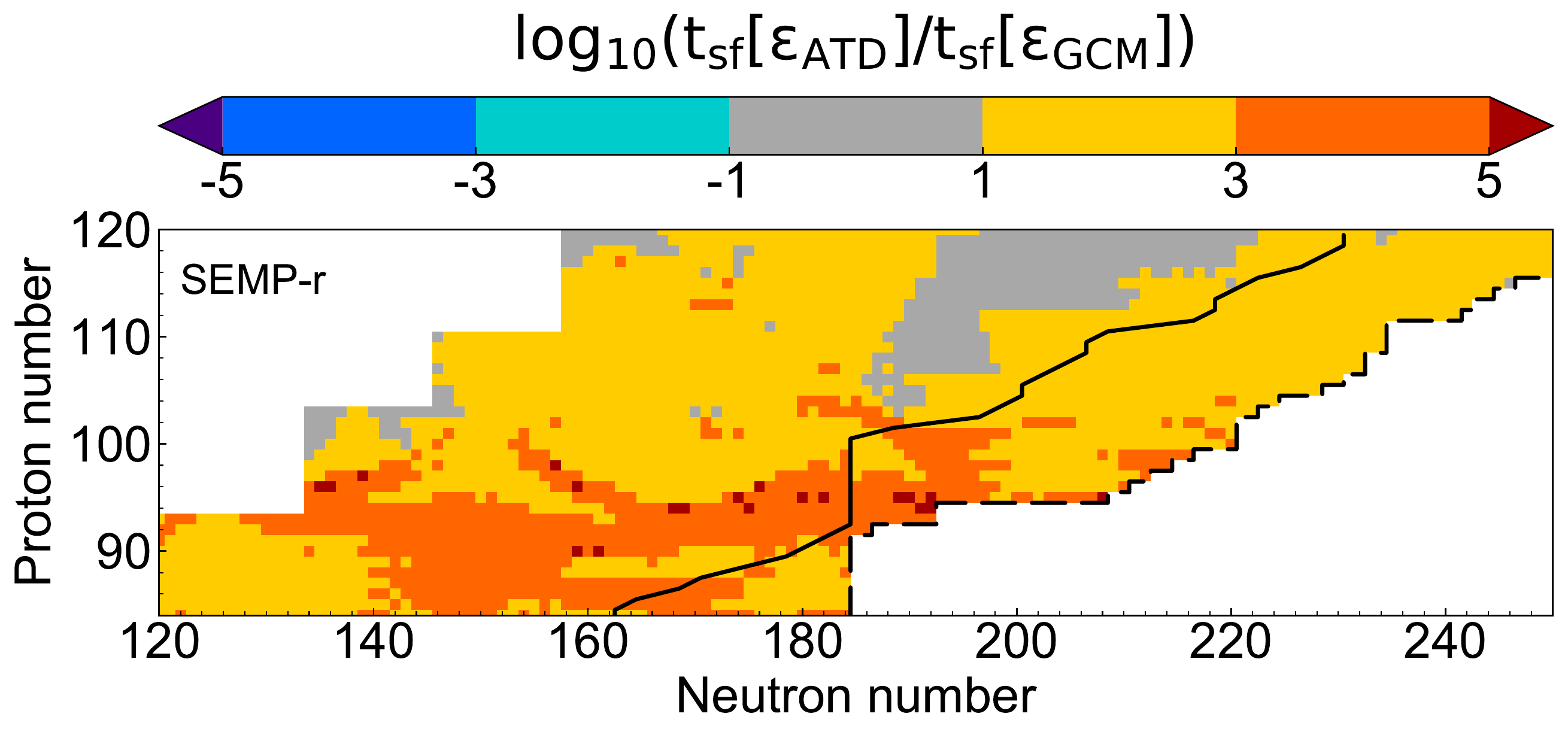}
 \caption{Sensitivity of the spontaneous fission lifetimes to 
 different values of the collective ground-state energy 
 $\log_{10}[t_\mathrm{sf}(\epsilon_\mathrm{vib}^\mathrm{ATDHFB})/%
 t_\mathrm{sf}(\epsilon_\mathrm{vib}^\mathrm{GCM})]$
 computed using the semiempirical collective inertias.}
 \label{fig:tsfZPE}
\end{figure}

For completeness we conclude this discussion studying the impact of the
vibrational zero-point energy correction $\ezpe$ on the spontaneous fission
lifetimes. Fig.~\ref{fig:tsfZPE} shows the logarithm of the ratio of $\tsf$
computed with the same semiempirical inertias and two different $\ezpe$
calculations, obtained from the ATDHFB and GCM formalisms of
Eqs.~(\ref{eq:ZPEA}) and~(\ref{eq:ZPEG}). We found that the $\tsf$ computed with
the ATDHFB $\ezpe$ are usually between 1 and 2 orders of magnitude larger than
the GCM ones for $r$-process nuclei. This variation is similar to the one
obtained for the collective inertias, in agreement with our conclusion that at
short timescales the $\tsf$ is more sensitive to the shape of the fission
barrier.

\subsection{$\alpha$-decay half-lives}\label{sec:alpha}
For completeness, we studied the competition between SF and $\alpha$ decay. The
$\alpha$-decay half-lives are obtained by means of the Viola-Seaborg
formula~\cite{Viola1966} using the recent parametrization of
Ref.~\cite{Dong2005}.  The main advantage of the Viola-Seaborg formula is that
it only requires the $Q_\alpha$ value of the parent nucleus to compute the
$\alpha$-decay half-lives.  Comparing our predictions with the AME2012 atomic
mass evaluation~\cite{Wang2012}, we found that BCPM reproduces the $Q_\alpha$
values with a rms deviation of 0.68~MeV.  For the $\alpha$-decay half-lives of
nuclei with $Z\ge84$, the logarithm of the mean and standard deviations are 1.92
and 2.51 respectively, corresponding to deviations between theoretical
half-lives and experimental data of factors 316 and 80. These large deviations
in the $t_\alpha$ reflect the difficulties of reaching accuracies beyond the
logarithmic precision in lifetimes calculations involving tunneling processes,
as it was already mentioned in Sec.~\ref{sec:bench}. Actually the accuracy of
the Viola-Seaborg formula itself is larger than a factor 6~\cite{Dong2005}, and
it is important to notice that the deviations obtained in the $t_\alpha$ are
smaller than those obtained in the $\tsf$ without renormalization of the
collective inertias.

\begin{figure}[tb]
 \includegraphics[width=0.49\textwidth]{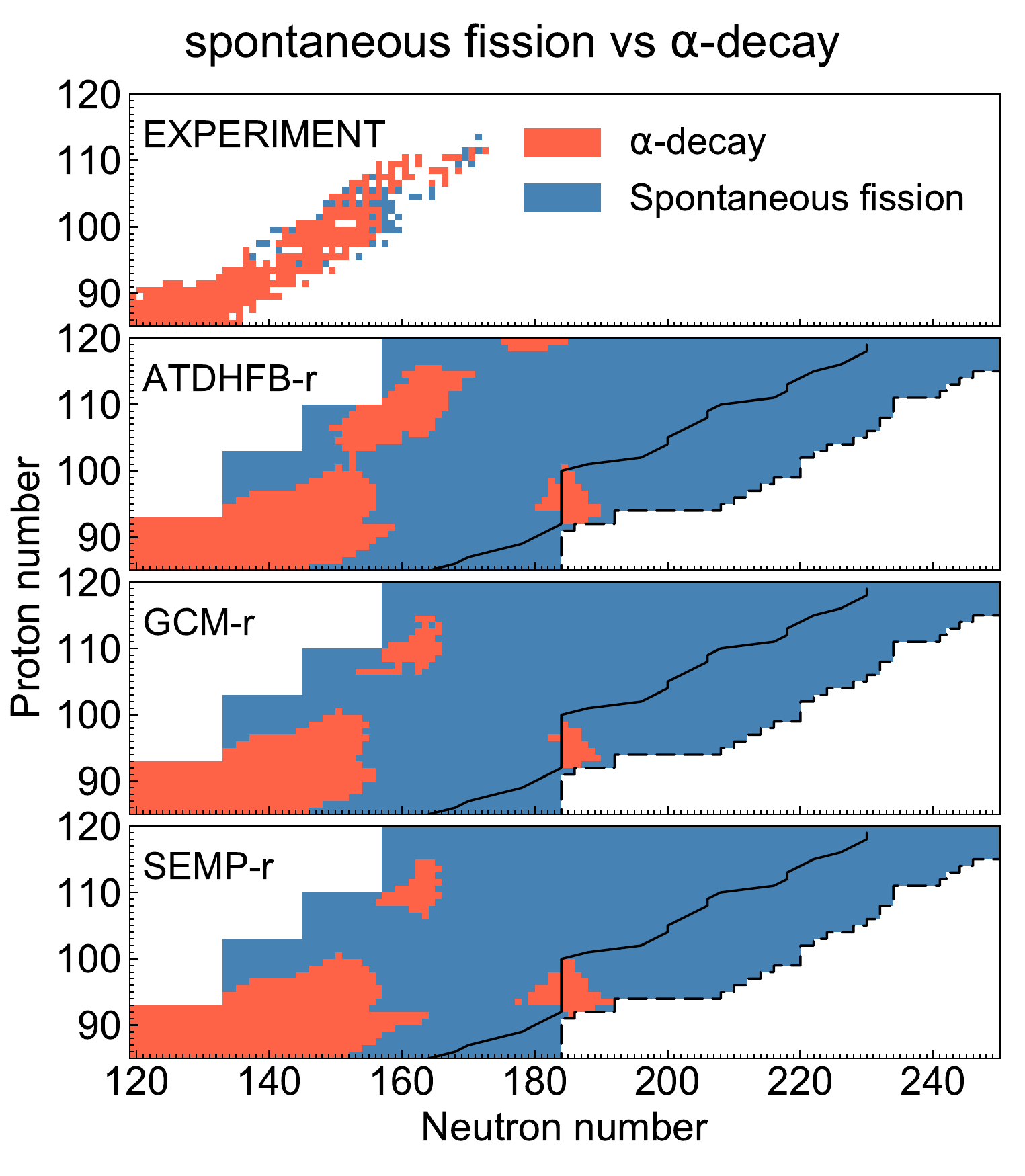}
 \caption{Dominating channel between $\alpha$ decay and spontaneous fission for 
 different collective  inertias. For  comparison, the upper panel shows the dominating 
 channel for nuclei experimentally observed~\cite{Wang2012}.}
 \label{fig:alpha}
\end{figure}

Fig.~\ref{fig:alpha} shows the dominating channel (either SF or $\alpha$ decay)
predicted by BCPM using different collective inertias, and the upper panel shows
the experimental data extracted from Ref.~\cite{Wang2012}. All the inertia
schemes predict $\alpha$ decay to be the dominating channel in the region $84\le
Z\le98$ and $118\le N\le156$, in good agreement with experimental data.  The
$\alpha$-decay lifetimes seem to be overestimated around $Z/N=102/150$
corresponding to a region of nuclei dominated by SF. One should also notice that
the jumps in the neutron separation energy around the predicted magic neutron
number $ N=184$ produces an island of nuclei dominated by $\alpha$-decay
centered on $Z=94$. Finally, the ATDHFB-r is the only scheme predicting the
$\alpha$-decay to be the dominating channel for nuclei in the island of
stability.

\subsection{Neutron induced rates}\label{sec:channels} 

We conclude our study showing the impact of collective inertias on the
calculations of rates relevant for $r$-process nucleosynthesis.  Following the
statistical picture described by the Hauser-Feshbach theory~\cite{Hauser1952},
for a target nucleus in the ground-state the neutron induced cross section for a
reaction $i^{\text{gs}}(n,\text{out})$ can be written as~\cite{Panov2009a}:
\begin{equation}
 	\begin{split}
 		\sigma_{(n,\text{out})}^{\text{gs}}=&\frac{\pi\hbar^{2}}{2 m
 		E_{\text{cm}}(2J_{i}^{\text{gs}}+1)(2J_{n}+1)}\\
 		&\sum_{J,\pi}(2J+1) \frac{T^{\text{gs}}_{n}(E,J^{\pi},E_i^\text{gs},J_i^\text{gs},\pi_i^\text{gs})
 		T_\text{out}(E,J^{\pi})}{T_\text{tot}(E,J^{\pi})}\,,
 	\end{split}
	\label{eq:cs}
\end{equation}
being $m$ the reduced mass, $E_{\text{cm}}$ the center-of-mass energy and
$T(E,J^{\pi})$ the transmission coefficient for a compound nucleus with energy
$E$, spin $J$ and parity $\pi$.
In the case of fission, the probability of the compound nucleus to
penetrate all the possible fission barriers is mimicked by adding
transition states on top of the saddle points. The fission
transmission coefficient $T_\text{fiss}$ is thus given
by~\cite{Goriely2009}:
\begin{equation}
	T_{\text{fiss}}(E,J^{\pi}) = \int_0^{E} d\varepsilon\,
	P(E,\varepsilon)\rho(\varepsilon,J^{\pi})\,,
	\label{eq:tfiss}
\end{equation}
with $P$ defined in equation~\eqref{eq:ptunneling}. 

\begin{figure}[tb]
 \includegraphics[width=0.49\textwidth]{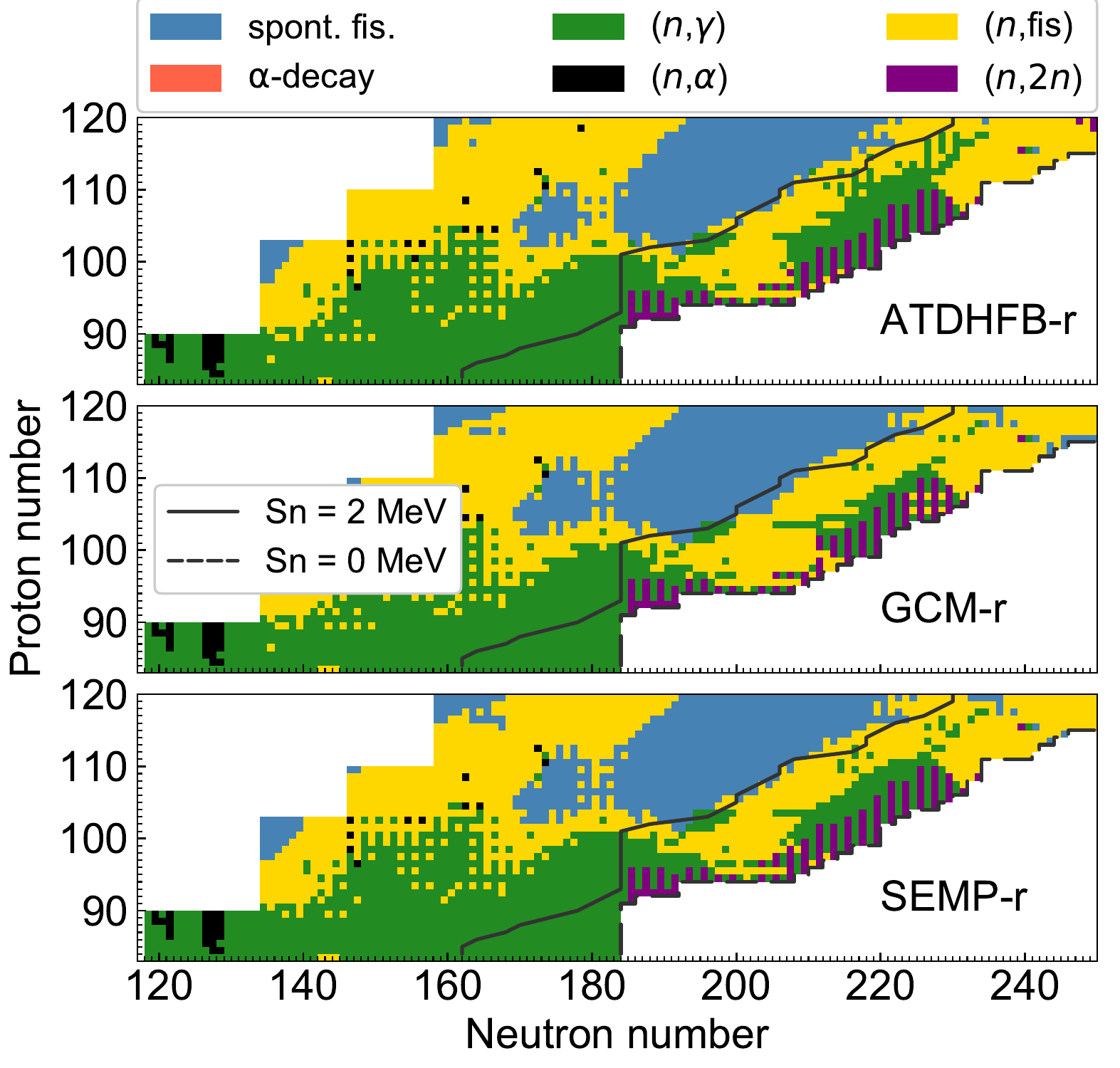}
 \caption{Dominating decay channels: spontaneous fission, $\alpha$-decay,
 neutron-capture, neutron-induced $\alpha$ emission, neutron-induced fission
 and two-neutron emission computed using different collective inertias
 schemes.}
 \label{fig:channels}
\end{figure}

Nuclei in astrophysical plasma exist in both ground and excited
states and the cross section of each excited state $\sigma^{x}$ 
contributes to the final stellar cross section $\sigma^{*}$ assuming that the
relative population of the states follows a Maxwell-Boltzmann distribution. The
stellar reaction rate for a given temperature $T$ is then calculated by folding
a Maxwell-Boltzmann distribution of relative velocities between projectiles and
targets with the stellar cross section~\cite{Fowler1974}:
\begin{equation}
	\begin{split}
		\langle \sigma v \rangle_{(n,\text{out})} =&
		\sqrt{\frac{8}{ \pi m (kT)^{3}}}\\
		&\int_0^{\infty}\sigma^{*}_{(n,\text{out})}(E)
		E\exp\left(-\frac{E}{kT}\right)dE\,.
	\end{split}
	\label{eq:sterate}
\end{equation}
Using Eq.~\eqref{eq:sterate} we computed the neutron capture,
neutron induced fission, neutron induced alpha decay and neutron
induced two-neutron emission stellar rates over the whole superheavy
landscape using the binding energies, fission barriers and the
renormalized collective inertias obtained from BCPM EDF. We adopted
the level densities given by the
constant temperature plus Fermi gas model and the Kopecky-Uhl
generalized Lorentzian gamma-ray strengths~\cite{Capote2009}.  The
calculations were carried out using the
TALYS~\footnote{\url{http://www.talys.eu/}} reaction code for
a range of temperatures between 0.01 and 10 GK~\cite{Koning:talys}. 
Fig.~\ref{fig:channels} shows
the dominating decay channel of each nucleus for typical conditions of
$r$-process in neutron star
mergers ($T=0.9\,\textrm{GK}$,
$n_n=1.0\times10^{28}\,\textrm{cm}^{-3}$)~\cite{Mendoza-Temis2014}.
Comparing the rates obtained from the different collective inertias we found
that all the schemes predict a very similar scenario with fission dominating
over the neutron capture for nuclei in the $r$-process path above the neutron
magic number $N=184$. There is a region above $N=206$ where neutron capture is
the dominating channel. However, the $r$ process path may never reach this
corridor due to the large $(n,\text{fis})$ and $(n,2n)$ rates at lower $N$. We
conclude then that independently of the computational scheme, the production of
nuclei heavier than $N>184$ will be strongly hindered due to the dominance of
neutron induced fission.

\section{Conclusions}\label{sec:conclusions} 

We have presented fission properties of 3640 superheavy
nuclei obtained within the Self-Consistent Mean-Field scheme and the
BCPM EDF.  The fission path is computed by minimizing the potential
energy using the axial quadrupole moment operator as a collective
degree of freedom and allowing for octupole and hexadecapole
deformations. The potential energy surface of nuclei with an odd
number of protons and/or neutrons is calculated using the PNAM,
maintaining the level of accuracy obtained for even-even nuclei. The
spontaneous fission lifetimes are evaluated using the WKB formula
involving the effective potential, collective inertias and collective
ground-state energy of the nucleus.  Both vibrational and rotational
corrections are properly subtracted from the effective potential.
Collective inertias are evaluated using three different schemes
(ATDHFB, GCM and the semiempirical formula) to test the sensitivity of
the spontaneous fission lifetimes. 

Comparing our results with the available experimental data we found that BCPM
tends to overestimate the spontaneous fission lifetimes, specially in the region
of light actinides where the fission barriers and collective inertias are
extremely large. In order to account for the effect of dynamic calculations in
the determination of the fission path, we propose a phenomenological approach
based on the renormalization of the collective inertias. By multiplying the
collective inertias by a constant factor the agreement with experiment can be
greatly improved. The comparison with experimental lifetimes showed that the
accuracy and precision of the theoretical predictions improve as the mass number
increases, providing the confidence to explore the region of nuclei relevant for
the $r$ process.

The landscape of the fission barrier obtained with the BCPM EDF show
five islands of local maxima.  Both the magic neutron numbers 126 and
184 lead to an increase of the fission barriers. Two other regions of
enhanced barriers are found corresponding to the peninsula and
island of stability. BCPM predicts a region of vanishing barriers for
nuclei with $ A \sim 292$, in agreement with other theoretical
models. By studying the energy window of the neutron induced fission,
we concluded that the $r$-process path may terminate in this region
and cycle to lighter fission products.

We performed a complete study of the sensitivity of the spontaneous fission
lifetimes to the quantities entering in the WKB formula. We studied the
variations on lifetimes when different schemes  of the collective inertias,
vibrational energy corrections and collective ground-state energies are used.
For $r$-process nuclei, the renormalized collective inertia schemes result in
variations of the spontaneous fission lifetimes smaller than three orders of
magnitude and only one order of magnitude when GCM and ATDHFB alone are
considered.
We also found that the spontaneous fission lifetimes are strongly affected by
the collective ground-state energy $E_0$. These large variations of the
lifetimes can be related to a complex structure of the fission paths where
several humps are present.  Some of the barriers and isomers are placed 0.6--1.2
MeV above the ground state and their contribution to the action integral
strongly depends on the value of $E_0$. Variations of the lifetimes on different
schemes of the vibrational corrections confirmed the major role played by the
fission barrier shape in the $\tsf$ of $r$ process nuclei.

Finally, we studied the competition of fission with other channels. We computed
the $\alpha$-decay half-lives by means of the Viola-Seaborg formula and compared
the results with the spontaneous fission lifetimes.  We found that in all the
schemes $\alpha$ decay dominates over spontaneous fission in nuclei with
$Z\le98$ in good agreement with experimental data. 
Using the statistical approach we computed neutron induced rates based on
binding energies, fission barriers and the non-renormalized collective inertias
obtained from the BCPM EDF. Our calculations showed that a proper treatment of
the inertias should lead to rates which are rather insensitive to the different
kind of inertias used, and that the synthesis of nuclei above $N>184$ during the
$r$ process will be strongly inhibited by the neutron induced fission.

The implementation of the BCPM rates in network calculations including a
consistent evaluation of beta delayed fission rates is necessary to determine
the impact on $r$-process nucleosynthesis. Work along these lines is already in
progress.

\begin{acknowledgments} 
  The authors are grateful to Hans Feldmeier for his helpful
  discussions on theoretical aspects of the fission process, to Stephane Goriely
  for providing the HFB14 fission barriers plotted in Fig.~\ref{fig:Bf-comp} and
  to Zachary Matheson for carefully reading this manuscript. SAG and GMP
  acknowledge support from the Deutsche Forschungsgemeinschaft through contract
  SFB~1245, and the BMBF-Verbundforschungsprojekt number 05P15RDFN1. The work of
  LMR has been supported in part by the Spanish MINECO Grants No.~FPA2012-34694
  and No.~FIS2012-34479 and by the ConsoliderIngenio 2010 Program MULTIDARK
  CSD2009-00064.
\end{acknowledgments}

\bibliography{library}
\end{document}